\documentstyle[preprint,aps,epsfig]{revtex} 
\newcommand \Pomeron {I\!\!P}
\begin{document}
\tighten
\title{COLOR COHERENT PHENOMENA ON NUCLEI AND THE QCD EVOLUTION EQUATION} 
\author{Leonid Frankfurt}
\address{Sackler School of Physics and Astronomy\\
Tel Aviv University, Ramat Aviv \\
Tel Aviv, Israel}
\author{Vadim Guzey}
\address{The Pennsylvania State University\\
University Park, PA, USA;\\
 Special Research Centre for the Subatomic Structure of Matter (CSSM),\\
 Adelaide University , Australia, 5005}
\author{and}
\author{Mark Strikman}
\address{The Pennsylvania State University\\
University Park, PA, USA}

\preprint{
\vbox{
\hbox{ADP-00-48/T429}
}}

\maketitle
\begin{abstract}

We review the phenomenon of color coherence in quantum chromodynamics (QCD), 
its implications 
for hard and soft processes with nuclei, and its experimental manifestations. The relation
 of factorization theorems in QCD with  
 color coherence phenomena 
in deep inelastic scattering (DIS) and color coherence phenomena  in hard exclusive 
processes is emphasized. Analyzing numerically  the QCD evolution equation for conventional and  
skewed parton densities in nuclei,
 we study the onset of generalized color transparency and
nuclear shadowing of the sea quark and gluon distributions in nuclei
 as well as related phenomena. Such novel results as the dependence of the effective coherence length on $Q^2$ and general trends of the
QCD evolution are discussed.
The limits of the applicability of the QCD evolution equation at small Bjorken $x$ 
are estimated 
by comparing the inelastic quark-antiquark- and two gluon-nucleon (nucleus) cross sections, calculated within the DGLAP approximation, with the dynamical boundaries, which follow    
from the unitarity of the $S$ matrix for  purely QCD interactions.
We also demonstrate that
 principles of color coherence  play an important role in the  
processes of soft diffraction off nuclei.

\end{abstract}

\section{Introduction}
\label{sec:Introduction}

The aim of this review article is to summarize fundamental principles of color coherent phenomena in quantum chromodynamics (QCD) and to apply these principles to the calculation of cross sections of
 diffractive processes at high energy, where color coherence  plays a crucial role. 

Our goal is threefold. Firstly, we make an attempt to give a comprehensive and pedagogical overview of main ideas which constitute the concept of color coherence in QCD. In the preparation of this part, we have benefited from review articles \cite{rev1,rev2}, a talk \cite{talk1}, and a series of lectures \cite{talk2}.

 Secondly, we discuss hard exclusive diffractive processes. 
The recent proof 
of the factorization theorem  for hard exclusive diffractive
electroproduction of mesons 
\cite{Collins} justifies  the computation of 
 the amplitude of 
this reaction in terms  of the skewed parton densities
 \cite{Collins,Radyushkin96}. 
 In the present work, we analyze numerically 
the onset of generalized color transparency in hard exclusive diffractive processes
and the phenomenon of nuclear shadowing in total cross sections in deep inelastic scattering (DIS) 
 using the QCD evolution equation for the skewed and conventional nuclear parton densities.
We also estimate the limits of the applicability of the 
DGLAP QCD evolution equation by comparing the calculated inelastic and elastic scattering cross sections of
spatially small quark and gluon dipoles off the nucleon or nucleus   
with the dynamical boundaries, which follow from
the unitarity of the $S$  matrix for purely QCD interactions \cite{FKS96}. 
We demonstrate that studies of a new QCD regime, color opacity, may be 
feasible with future electron-nucleus colliders 
and in special hard processes in $ep$ collisions at HERA.

Thirdly, we review 
the soft diffraction of hadrons on light and heavy nuclei. We demonstrate that the principles of color coherence form the ground for the adequate phenomenological description of the available data on hadron-nucleus coherent diffractive dissociation.  
  
The phenomenon of color coherence in QCD has two principal features: 1) an energetic projectile (hadron or photon) consists of coherent quark-gluon configurations of very different transverse spatial sizes; 2) a small-size  quark-gluon configuration has reduced interactions with the target
at not very small $x$.

The coherent quark-gluon configurations, which constitute the energetic projectile, are called color fluctuations because  QCD color occupies different volumes for the configurations of different transverse sizes.
 Phenomenon 2) is  called color screening. Both aspects of color coherence have been proven in perturbative QCD (pQCD) and have been observed experimentally.
Since configurations of various transverse sizes are present in the projectile at high energies, studies of color coherence phenomena provide an important relation between hard physics (small-size quark-gluon fluctuations), which is quite well understood in terms of QCD,  and soft physics (large-size quark-gluon fluctuations), which is treated mostly phenomenologically.

Property 1 is based on the observation that at high energies the longitudinal distance, essential for a given process,  increases with the increase of energy. Originally,
 before QCD was known,
such a behaviour of high-energy hadron collisions   was suggested in \cite{Gribov1,Ioffe68,Gribov2,Gribov3}. 
The longitudinal distance is called  coherence length $l_{c}$ \cite{FS91}.
The word {\it coherence} reflects the attribute  of any quantum field theory, 
which consists in the fact that particles are
predominantly radiated at large and increasing with energy distances
because  of Lorentz slowing down of interactions.

Property 1  is well understood in quantum electrodynamics \cite{LP53}.
Coherence length $l_c$ 
also exists in pQCD where it also   
increases  with energy. It can be easily proven that this  property of $l_{c}$ is a consequence
of gauge invariance and the renormalizability of QCD.
The convergence of integrals over ``masses'' of the intermediate states in the wave function of the incoming virtual photon 
follows from the absence of ultraviolet divergencies in the 
amplitudes, except where it is related to 
the 
renormalization of the coupling constant. However, in QCD this convergence 
is rather  slow, which leads to the $Q^2$ dependence of the coherent phenomena,
discussed in this review. This $Q^2$
dependence 
 can also be understood as a result of the slow $Q^2$ dependence
of the effective coherence length, introduced in the present paper.

 The large coherence length allows  the presence of many quark-gluon coherent fluctuations of various transverse sizes in the energetic projectile. The latter results in a large probability of coherent diffractive processes,
experimentally  observed at HERA \cite{HERA},
 and  leading twist nuclear shadowing in DIS, experimentally  observed 
for light nuclei \cite{CERN1} and heavy ones \cite{CERN2}
at CERN. Below we shall consider the relationship between the large $l_{c}$ and the experimental results 
 in more detail.

Let us consider diffractive DIS scattering of a photon with
virtuality $Q^2$, energy $q_{0}$, and high momentum $q$
 on a hadronic target $T$ in the reaction $\gamma^{\ast}+T \to X+T^{\prime}$. 
The coherence length $l_{c}$ is introduced as the time (distance) for which the 
incoming virtual photon exists as quark-gluon 
fluctuations with the average mass squared  $\langle M_{X}^2 \rangle$.
It follows from the uncertainty principle that
\begin{equation}
l_{c}=\frac{1}{\Delta E}=\frac{1}{\sqrt{q^2+\langle M^2_{X} \rangle }-q_{0}} \simeq \frac{2q}{\langle M^2_{X} \rangle +Q^2} \ ,
\label{cl1}
\end{equation}  
where $m_{N}$ is the nucleon mass. 
In  Eq.\ (\ref{cl1}) one often uses the approximation $\langle M^2_{X} \rangle = Q^2$ and arrives at
\begin{equation}
l_{c} \simeq \frac{1}{2 m_{N} x} \ ,
\label{cl1b}
\end{equation}
where $x$ is the Bjorken variable \cite{Ioffe68}.
 In Subsect.\ \ref{subsec:LC} we shall explain that 
since coherence of high energy processes is realized through 
diffractive processes, the parton model
approximation $\langle M^2_{X} \rangle = Q^2$ is no longer valid if one takes into account  
the effects of the QCD evolution. 
Then, more appropriate quantities are  the effective coherence length 
$l_{c}^{eff}$ defined by Eq.\ (\ref{cl1}), where $M^2_{X}$ is 
the typical mass in diffractive processes in DIS.

Let the coherence length of Eq.\ (\ref{cl1}) be larger than the typical size of the target  
\begin{equation}
l_{c} > 2R_{T} \ ,
\label{cond} 
\end{equation}  
where $R_{T}$ is the  radius of the target.
Then, not only are all quark-gluon fluctuations, satisfying Eq.\ (\ref{cond}),  present in  the photon wave function and  interact coherently with the target $T$,
 but  the target also has a large probability of remaining  in the ground state.
 This fact leads to a large probability of coherent (when the nucleon or nuclear target  is left intact during the interaction) diffractive processes on nucleons and nuclei.

In inclusive DIS on nuclei, when Eq.\ (\ref{cond}) holds, the incoming virtual photon  interacts coherently with several nucleons of the nucleus. This manifests itself in nuclear shadowing.  The QCD-improved aligned jet model, which takes into account the interplay between soft and hard physics \cite{FS1},
 explicitly demonstrates that  shadowing should be of the leading twist, i.e. only the logarithmic decrease of the effect of shadowing with increasing $Q^2$  is allowed.  
Note that at sufficiently small $x$ the very concept of 
decomposition over twists 
might appear inapplicable to QCD. We refer the reader to Subsect.\ \ref{subsec:Unitarity}  for the relevant discussion.

Property 2) of color coherence, color screening, 
has been proved 
in pQCD using gauge invariance, 
asymptotic freedom,  and the QCD  factorization theorem.  
The suppression of the interaction of small-size 
quark-gluon configurations of incoming hadrons and 
photons with hadronic targets
leads to several 
distinctive
 experimental signatures of 
color coherence. In nuclear quasi-elastic reactions, where 
small-size configurations are produced as a result 
of high $Q^2$ two-body wide-angle scattering, the 
suppression of initial and final state interactions is 
called color transparency \cite{Mueller82,Brodsky82}.
In QED, a similar phenomenon of ``the electric neutrality'' was theoretically predicted in
as early as 1954 by H.~Bethe and L.~Maximon \cite{Bethe} by considering coherent
production of 
$e^+e^-$ pairs in photon-nucleus scattering
 within  the eikonal approximation\footnote{We are indebted
 to S.~Brodsky who drew our attention to \cite{Bethe}.}.
Note that one can justify the applicability of
the eikonal approximation in QED, while, in QCD, the validity of
this approximation is questionable 
even in the perturbative regime because  of ``strong''
gluon-gluon interactions.
In hard diffractive 
processes, like 
the electroproduction of vector mesons
off nucleons and nuclei,
the interaction becomes strong 
at sufficiently small $x$, $x \leq 0.01$,
 and one reaches the regime of perturbative color
 opacity \cite{Brod94,FMS93}.
In this regime,
the suppression in the amplitude is explained by nuclear shadowing of gluons and, hence, 
the factorization theorem is
still valid. 
Therefore, 
the regime of perturbative color
 opacity
 can be also called the regime of  
generalized color transparency. We demonstrate that, in 
the limit of fixed $x$ and  $Q^2\to \infty $, 
the generalized color transparency is reduced to color transparency.
At sufficiently small $x$, the interaction becomes strong, which leads to violation of the factorization theorem,
and, possibly, to different nuclear shadowing in hard inclusive and hard exclusive
processes.

Perturbative QCD  
causes a particular energy and $Q^2$ behaviour of the 
differential cross section of the hard vector meson production.
 This  behaviour has been verified experimentally (see discussion of  relevant 
HERA experiments in \cite{HERA}).
Color transparency has recently been observed directly 
in the diffractive excitation of pions into two jets
off nuclear targets \cite{Ashery}, in line with the predictions of \cite{FMS93}.
A number of other processes such as $J/\psi$ photoproduction, 
leading hadron production in nuclear reactions using 
Drell-Yan lepton pairs 
 as a trigger, and the production of leading nucleons in nucleus-nucleus collisions can be understood and confronted with the experimental data only if the coherent nature of the reactions is taken into account.

This review  is organized in the following way. Sect.\ \ref{sec:CC} is concerned with cornerstones of the theoretical treatment of color coherent phenomena in QCD. These are the dynamics of small objects and the factorization theorems. We also discuss 
the relationship between diffraction on the proton in the reaction $\gamma^{\ast}+p \to X +p^{\prime}$ and nuclear shadowing in DIS scattering $\gamma^{\ast}+A \to X$.
We critically examine the definition of the coherence length and present the  $Q^2$ dependence of the effective coherence length $l_{c}^{eff}$.  
We also review the  aligned jet model of Bjorken and explain the improvements of this model which are  to be done to make it conform with QCD \cite{FS1}.  

In Sect.\ \ref{sec:HD}, the principles of color coherence in QCD are applied  to exclusive hard diffractive processes. 
The main features of generalized color transparency are reviewed.
We also discuss  experiments, in which  color coherence plays a crucial role. New reactions, where the two-gluon exchange is prohibited but color transparency still holds, are proposed.

Sect.\ \ref{sec:NA} quantitatively illustrates our conclusions concerning 
the transition from generalized color transparency to color transparency
at fixed $x$ and $Q^2\to \infty$, and concerning
 the behaviour of parton densities in nuclei
by the virtue of analyzing the QCD evolution of skewed and conventional  nuclear parton densities. A study of general trends of the QCD evolution in the $x-Q$-plane is presented. 
We also study  the  limits of  applicability of the leading twist DGLAP evolution equation at small Bjorken $x$
using the unitarity of the $S$-matrix for purely QCD interactions as a  guiding principle \cite{FKS96}.

 In Sect.\ \ref{sec:SD}, we give an overview of soft hadron diffraction on nuclei from the point of view of color coherence. We demonstrate that the formalism, which  takes into account color fluctuations of hadronic projectiles explicitly, gives  
a universal description of the available data on diffractive dissociation and nuclear shadowing for light and heavy nuclei.

\section{Color Coherence in diffractive processes}
\label{sec:CC}

The investigation of the role of color coherent phenomena in inclusive, exclusive and semi-exclusive high energy processes is one of the rapidly developing fields in QCD. Many calculations can be done in pQCD with the use of the dynamics of small-size quark-gluon configurations and the factorization theorems in a model-independent way.

 We begin this section by reviewing a deep relationship 
between nuclear shadowing in inclusive DIS  on nuclei and 
DIS diffraction on protons, which is based on the factorization theorem for DIS diffraction $\gamma^{\ast}+p \to X +p^{\prime}$ \cite{Collins2}.

\subsection{Leading twist nuclear shadowing and diffraction in DIS}
\label{subsec:shadowing}

A deep connection between high energy diffraction and the phenomenon of nuclear shadowing was first understood by V.~Gribov \cite{Gribov2,Gribov4}. On the qualitative level, nuclear shadowing in DIS on the lightest nucleus, deuterium, arises from the  interference between the amplitudes for diffractive scattering of the projectile  off the proton and the neutron of the deuterium target. This is presented schematically in Fig.\ \ref{fig:str1}. The application of the cutting rules of  Abramovski$\breve{{\rm i}}$, Gribov, and Kancheli  \cite{AGK} explicitly demonstrates that the interference diagram  decreases the total cross section for $\gamma^{\ast}D$ scattering, i.e. it leads to nuclear shadowing. 

One can use this result in order to calculate the modification of parton densities in nuclei at low values of Bjorken $x$. By comparing the QCD diagrams for hard diffraction and for nuclear shadowing due to scattering off two nucleons (see Fig.\ \ref{fig:str1}), one can prove \cite{FS99} that, in the limit of low nuclear thickness, nuclear shadowing of nuclear parton densities $f_{j/A}(x,Q^2)$ is  unambiguously expressed through diffractive parton densities in the proton $f^{D}_{j/N}(\beta,Q^2,x_{\Pomeron},t)$
\begin{eqnarray}
&&f_{j/A}(x,Q^2)/A  =  f_{j/N}(x,Q^2)-{1 \over 2}\int d^2b\int_{-\infty}^{\infty}dz_1\int_{z_1}^{\infty} dz_2 \int_x^{x_0} dx_{\Pomeron}  \nonumber \\
&&\times \frac{1-\eta^2}{1+\eta^2}\, f^{D}_{j/N}\left(\beta, Q^2,x_{\Pomeron},t\right)_{\left|k_t^2=0\right.}
\rho_A(b,z_1)\rho_A(b,z_2) \cos(x_{\Pomeron}m_N(z_1-z_2)) \ ,
\label{sh1}
\end{eqnarray}
where $f_{j/N}(x,Q^2)$ is the conventional parton density in the proton; $\rho_A(r)$ is the nucleon density in the nucleus with atomic number $A$; $x_{\Pomeron}$ is the fraction of the proton longitudinal momentum carried by the Pomeron; $t$ is the momentum transfer, defined as $-t=(k_{t}^2+(x_{\Pomeron} m_{N})^2)/(1-x_{\Pomeron})$, where $k_{t}$ is the transverse component of the momentum, transferred to  the struck nucleon; $\beta=x/x_{\Pomeron}$; $x_{0} \simeq 0.02$. In Eq.\ (\ref{sh1}), the factor $(1-\eta^2)/(1+\eta^2)$, where $\eta=-\pi / 2\, \partial \ln(\sqrt{f^D_{i/N}})/ \partial  \ln(1/x_{\Pomeron})=\pi / 2 \, \kappa$ 
 accounts for
the real part of  the amplitude for the diffractive scattering \cite{AFS99}. The experimental value of the parameter $\kappa$, defined as $f^{D}_{j/N}\left(\beta,Q^2,x_{\Pomeron},t\right)_{\left|k_t^2=0\right.}
\propto x_{\Pomeron}^{2\kappa}$, is rather small,  $\kappa\approx 0.15 $.

Considering deuteron, one can write a compact expression, which  includes the
$t$ dependence of the electromagnetic form factor $F_{^2{\rm H}}(q)$, where $-t=q^2$ 
\begin{eqnarray}         
f_{j/^2{\rm H}}(x,Q^2)/2  =  f_{j/N}(x,Q^2)-{1 \over 8\pi}
\int_{x}^{x_{0}} dx_{\Pomeron} d^2 t
\frac{1-\eta^2}{1+\eta^2}\, f^{D}_{j/N}\left(\beta, Q^2,x_{\Pomeron},t\right)F_{^2{\rm H}}(4t) \ .
\label{shdeu}
\end{eqnarray}
Similarly to Eq.\ (\ref{shdeu}), one can easily generalize 
Eq.\ (\ref{sh1}) in order to include the dependence of the diffractive amplitude on $t$ ( see, e.g., \cite{LS}).

The factorization theorem for DIS diffraction  $\gamma^{\ast}+p \to X +p^{\prime}$ \cite{Collins2} states that the diffractive parton densities at fixed $x_{\Pomeron}$ and $t$ are described by the leading twist DGLAP QCD evolution equation. Therefore, the nuclear shadowing, described by Eq.\ (\ref{sh1}), is also a  leading twist effect and is governed by the DGLAP QCD evolution equation \cite{FS99}. 

Thus, Eq.\ (\ref{sh1}) should be understood as the initial condition for the QCD evolution for the nuclear parton densities. Since  Eq.\ (\ref{sh1}) is valid at low $x$, in particular at $l_{c}^{eff} > 2 R_{A}$, where $R_{A}$ is the nuclear radius (see Eq.\ (\ref{cond})), the nuclear effects typical for larger $x$ are absent
in Eqs.\ (\ref{sh1}) and (\ref{shdeu}).
 A slight enhancement of gluon and valence quark parton densities in the nucleus at $x \approx 0.1$ should be added to  Eqs.\ (\ref{sh1}) and (\ref{shdeu}) at some initial evolution scale $Q_{0}$. At larger $Q$, this enhancement will contribute at much smaller $x$ and will diminish the effect of nuclear shadowing. This simple argument shows that nuclear shadowing should decrease 
logarithmically as $Q$ increases due to the QCD evolution due to two effects:
 decreasing $l_c^{eff}$ with  increasing $Q^2$ and the growing 
contribution of the enhancement region $x\geq 0.1$. 

Eq.\ (\ref{sh1}) can be used to study 
 the amount of nuclear shadowing for quark and gluon parton densities
separately. The analysis of \cite{FS99} showed that: i) there is  significant shadowing for quark parton densities with small theoretical uncertainties; ii) nuclear shadowing for gluon parton densities is larger by approximately  a factor of 3 than for the quark parton densities\footnote{Note that this result was derived in the approximation of the low nuclear thickness. Hence,  it is valid for  light nuclei only.
For heavy nuclei, it is necessary to account for the interaction
of $\gamma^{\ast}$ with 3, 4, $\dots$ nucleons of the target.
Also, we have checked that the use of the diffractive parton densities obtained by Hautmann, Kunszt and Soper, based on the calculations in perturbative QCD at the initial scale $Q_{0}$=1.5 GeV \cite{HKS}, lead to very similar numerical results.}.  

Fig.\ \ref{fig:shh} presents the amount of nuclear shadowing at small $x$ for the ratio of nuclear and nucleon gluon parton densities $G_{A}/AG_{N}$ and for the ratio of nuclear and nucleon quark parton densities $q_{A}/Aq_{N}$ for the nuclei of lead (Pb) and carbon (C). At the initial scale of the QCD evolution $Q_{0}$=2 GeV, the nuclear shadowing corrections were calculated using Eq.\ (\ref{sh1}) as well as taking into account  the higher order rescattering terms. 
The dashed, dotted and solid curves  correspond to $Q$=2, 5 and 10 GeV, respectively.   One can see from Fig.\ \ref{fig:shh} that at the initial QCD evolution scale $Q$=2 GeV, nuclear shadowing for gluons is larger by approximately a factor of 3 than for quarks. As explained above, due to the QCD evolution, nuclear shadowing for the ratio $G_{A}/AG_{N}$  becomes significantly reduced at $Q$=5 and 10 GeV and is compatible to nuclear shadowing for  
$q_{A}/Aq_{N}$.

Note that, in  many models of nuclear shadowing which 
do not use the information about gluon-induced 
diffraction, it is expected that gluons are shadowed less than quarks.
Also, in a number of models, the interaction of small dipoles with 
several nucleons is considered in the eikonal approximation. 
However, 
by comparing Eq.\ (\ref{sh1}) and the expression based on the 
eikonalization of the pQCD cross section for the interaction of 
a small-size dipole with a
single nucleon (Eq.\ (\ref{cs})), it is easy to check
that the leading twist coupling of the small-size dipole to two nucleons, presented 
in Fig.\ \ref{fig:str1}, gives significantly larger shadowing than the 
higher twist 
eikonal diagrams, where four gluons are attached to a $q\bar q$ dipole.
For example, at $Q^2$=10 GeV$^2$ and $x=10^{-3}$, nuclear shadowing, calculated using  
Eq.\ (\ref{sh1}), is larger by a factor 3 than nuclear shadowing, calculated in the eikonal approximation.
Hence, the eikonal approximation strongly underestimates 
the strength of nuclear shadowing
for the interaction of small dipoles.
In particular, for the total photoabsorption cross section of longitudinally polarized virtual photons $\sigma_L$,  Eq.\ (\ref{sh1}) corresponds to the
 leading twist shadowing for gluons, while calculations, using the eikonal approximation, predict higher twist  nuclear 
shadowing, which  is much smaller numerically, cf.\ \cite{FKS96}.

At the same time, it was recently shown \cite{Capella} that a model, which incorporated 
the Gribov theory of nuclear shadowing and HERA diffractive data, successfully described the high precision NMC data on nuclear structure functions at low $x$ \cite{CERN1,CERN2}.

In the laboratory reference frame, nuclear shadowing is controlled by Eq.\ (\ref{sh1}) and by the coherence length $l_{c}$. In the next subsection we shall review the definition of $l_{c}$ 
and analyze its $Q^2$ dependence within the framework of the factorization theorem \cite{Collins2} and the QCD evolution equation.

\subsection{$Q^2$ dependence of the effective coherence length}
\label{subsec:LC}

It is evident from  Eq.\ (\ref{sh1}) that the leading 
contribution to nuclear shadowing in DIS on 
nuclear targets is proportional to the diffractive 
differential cross section at the minimal momentum transfer $t_{min}$.
Also, diffraction plays a dominant role in the generalized color 
transparency phenomenon. Thus, it is useful to introduce 
the notion of the effective coherence length $l_{c}^{eff}$.
By definition  $l_{c}^{eff}$
is the coherence length, which controls the onset of the nuclear 
shadowing and generalized color transparency phenomena.
The definition of the coherence length $l_{c}$, given 
by Eq.\ (\ref{cl1b}), is based on the parton model assumption that 
$\langle M_{X}^2 \rangle  \approx Q^2$. This 
approximation is valid only at moderate  $Q^2$
and $x\sim 10^{-2} \div 10^{-3}$.
 As 
$Q^2$ increases, the ratio $\langle M_{X}^2 \rangle /  Q^2$
increases 
logarithmically at fixed  $x$ 
 due to the QCD evolution. This effect becomes even more important at very small $x$
due to the presence of $\alpha_s\ln(1/x)$ effects.

According to the discussion in the Introduction, $\langle M_{X}^2 \rangle$ 
should be understood as 
the average mass squared of the 
diffractively produced final state $X$ in  
the single-diffractive scattering process  
 $\gamma^{\ast}+T \to X + T^{\prime}$. Thus, in order to study the $Q^2$ dependence of $l_{c}^{eff}$, 
one should determine  the typical masses  $\langle M_{X}^2 \rangle$, which give the dominant contribution to the diffractive differential cross section $d \sigma^{D} /dt(t=t_{min})$ of the process  
$\gamma^{\ast}+T \to X + T^{\prime}$, where 
$-t_{min} \approx (x_{\Pomeron}m_{N})^2$.

The diffractive differential cross section $d \sigma^{D} /dt(t=t_{min})$ 
can be written as
\begin{equation}
\frac{d \sigma^{D}}{dt}(t=t_{min}) \equiv \frac{d^3 \sigma^{D}}{dx\,dQ^2\,dt}(t=t_{min})=\frac{2 \pi \alpha^2}{x Q^4}\Big(1+(1-y)^2\Big) \int_{x}^{0.02} dx_{\Pomeron}\, F_{2}^{D(4)}(x,Q^2,x_{\Pomeron},t_{min}) \ , 
\label{dcs1}
\end{equation}
where $F_{2}^{D(4)}(x,Q^2,x_{\Pomeron},t)$ is the diffractive structure
 function; $y$ is the rapidity. For 
simplicity, the longitudinal structure function is neglected in Eq.\ (\ref{dcs1}). 
The Pomeron model of diffractive processes is supposed to be valid at small $x_{\Pomeron}$. We followed the prescription of Ref.\ \cite{FS99} and set the upper limit of integration in Eq.\ (\ref{dcs1}) to 0.02.  
However, our numerical analysis is not too sensitive to the particular choice of the upper limit -- a variation of the upper limit between 0.01 and 0.03 changes the result for $l_{c}^{eff}$ by less than 25\%.

 The factorization theorem for diffractive DIS scattering \cite{Collins2}
states that one can introduce the diffractive parton
densities for fixed $x_{\Pomeron}$ and $t$, which satisfy the usual DGLAP evolution equation. The leading twist contribution to
$F_{2}^{D(4)}(x,Q^2,x_{\Pomeron},t)$
can be expressed through these parton densities.
In practical applications, a further simplifying assumption is made -- for sufficiently small $x_{\Pomeron} \leq 0.02$, where 
the vacuum 
pole
exchange dominates, one can write 
$F_{2}^{D(4)}(x,Q^2,x_{\Pomeron},t)$
in the factorized form
\begin{equation}
F_{2}^{D(4)}(x,Q^2,x_{\Pomeron},t)=f_{\Pomeron/p}(x_{\Pomeron},t) F_{2}^{D(2)}(\beta,Q^2) \ ,
\label{factoriz}
\end{equation}
where $f_{\Pomeron/p}(x_{\Pomeron},t)$ is the Pomeron flux; $F_{2}^{D(2)}(\beta,Q^2)$ is the diffractive structure function; $\beta$ is the fraction of the Pomeron momentum carried by  partons inside the Pomeron defined as $\beta=x/x_{\Pomeron}$. 
Substituting Eq.\ (\ref{factoriz}) in Eq.\ (\ref{dcs1}), one obtains
\begin{eqnarray}
\frac{d \sigma^{D}}{dt}(t=t_{min})=&&\frac{2 \pi \alpha^2}{x Q^4}\Big(1+(1-y)^2\big) \int_{x}^{0.02} dx_{\Pomeron}\, f_{\Pomeron/p}(x_{\Pomeron},t_{min}) F_{2}^{D(2)}(\frac{x}{x_{\Pomeron}},Q^2) \nonumber\\
=&& \frac{2 \pi \alpha^2}{x Q^4}\Big(1+(1-y)^2\big) \int^{1}_{x/0.02}\frac{d \beta\,x}{\beta^2} f_{\Pomeron/p}(\frac{x}{\beta},t_{min}) F_{2}^{D(2)}(\beta,Q^2) \ .
\label{dcs2}
\end{eqnarray}

In order to use Eq.\ (\ref{dcs2}) for studying $l_{c}^{eff}$ and  $M_{X}^2$  as a function of $Q^2$, let us notice that $M_{X}^2$ is kinematically related to $\beta$ as  
\begin{equation}
\beta=\frac{Q^2}{Q^2+M_{X}^2} \ .
\label{beta}
\end{equation}
This equation demonstrates that,  at fixed by the kinematics of the reaction $Q^2$ and $x$, the averaging over diffractive masses $M_{X}$ in Eq.\ (\ref{cl1})  is equivalent to the averaging over $\beta$.
Substituting Eq.\ (\ref{beta}) into  (\ref{cl1}) we obtain
\begin{equation}
l_{c}^{eff}=\frac{1}{m_{N}x}\langle \beta \rangle(Q^2,x) \ .
\label{cl3}
\end{equation}
$\langle \beta \rangle(Q^2,x)$  is the typical value of $\beta$, which dominates the integral in Eq.\ (\ref{dcs2}). It depends on  $Q^2$ and $x$. The $Q^2$ dependence of $\langle \beta \rangle(Q^2,x)$ is governed by the DGLAP QCD evolution of the diffractive parton densities, which enter $F_{2}^{D(2)}(\beta,Q^2)$. The $x$ dependence of $\langle \beta \rangle(Q^2,x)$ comes from the $x$ dependence of the lower limit of integration and the Pomeron flux in Eq.\ (\ref{dcs2}). 

We propose the following ansatz for the evaluation of  $\langle \beta \rangle(Q^2,x)$. At each $x$ and $Q$, $\langle \beta \rangle(Q^2,x)$ is given by the median of the integral in Eq.\ (\ref{dcs2})
\begin{equation}
\frac{1}{2}\int^{1}_{x/0.02}\frac{d \beta}{\beta^2} f_{P/p}(\frac{x}{\beta},t_{min}) F_{2}^{D(2)}(\beta,Q^2)=\int^{\langle \beta \rangle(Q^2,x)}_{x/0.02}\frac{d \beta}{\beta^2} f_{P/p}(\frac{x}{\beta},t_{min}) F_{2}^{D(2)}(\beta,Q^2) \ .
\label{beta2}
\end{equation}
Thus, solving Eq.\ (\ref{beta2}) for $\langle \beta \rangle(Q^2,x)$ will give the desired $Q^2$ dependence of $l_{c}^{eff}$ through Eq.\ (\ref{cl3}). 

In our analysis of Eq.\ (\ref{beta2}), we used the leading order expression for the diffractive structure function $F_{2}^{D(2)}(\beta,Q^2)$
\begin{equation}
F_{2}^{D(2)}(\beta,Q^2)=\sum_{a} e_{a}^2\, \beta \, q_{a}(\beta, Q^2) \ ,
\end{equation}
where $q_{a}(\beta, Q^2)$ is the parton density of a quark with flavor $a$ inside the Pomeron; $e_{a}$ is the charge of the quark. 

The $Q^2$ dependence of $q_{a}(\beta, Q^2)$ was studied  using the next-to-leading order (NLO) DGLAP QCD evolution equation. The recent parameterization of diffractive parton densities of Ref.\ \cite{ACW} was used 
as  an  input for the QCD evolution.
 We choose the fit {\bf D}, which gives the best description of the diffractive DIS  and photoproduction data from HERA.
All fits of  \cite{ACW} correspond to the initial scale $Q_{0}$=2 GeV. At this scale,  the gluon density is non-zero, $u(x)=d(x)=\bar{u}(x)=\bar{d}(x) \neq 0$, and $s(x)=\bar{s}(x)=0$. As one performs the QCD evolution to larger $Q > Q_{0}$, the equality $u(x)=d(x)=\bar{u}(x)=\bar{d}(x)$ remains and  strange quarks  $s(x)=\bar{s}(x) \neq 0$ are generated.

The current HERA data does not cover the region of small $\beta$ at small $x$, 
hence the fits of
\cite{ACW} are not sensitive to the region
where the triple Pomeron contribution dominates. In order to explore the sensitivity
of the value of $l_{c}^{eff}$ to 
this region, we have included the small-$\beta$ contribution, using the limiting
fragmentation ansatz of Ref.\ \cite{FKSpi}.
In the limit of small $\beta$ and at $Q^2=Q_0^2$, 
the multiplicity of leading hadrons
in the target fragmentation
region approximately coincides with that in photoproduction. 
This prediction is consistent with recent HERA data at $x_{\Pomeron}\geq 0.05$.

In Fig.\ \ref{fig:lqk} we present $l_{c}^{eff}m_{N}x
=\langle \beta \rangle(Q^2,x)$ as a function of $Q$ at $x$=10$^{-4}$, 10$^{-3}$, 
and 10$^{-2}$, calculated using Eqs.\ (\ref{cl3}) and (\ref{beta2}) 
with (dashed lines) and without (solid lines) small-$\beta $ contribution. 
 One can see that, at each $x$, the coherence length $l_{c}^{eff}$ is 
a decreasing function of $Q$. Also, one observes that the small-$\beta $ tail becomes
increasingly important with decreasing  $x$, especially at $Q^2 \sim Q_0^2$.

Using  Fig.\ \ref{fig:lqk}, one can also study the $x$ dependence of $l_{c}^{eff}$ at given $Q^2$. While $l_{c}^{eff}$ increases as $x$ decreases, this increase is slower than $1/x$ behaviour predicted by Eq.\ (\ref{cl1b}).
Also, the difference between $l_{c}^{eff}$ for different $x$ becomes larger as $Q^2$ increases.
 These effects are attributed to the QCD evolution of diffractive parton densities.

Note that the definition of $l_{c}^{eff}$, which  uses Eqs.\ (\ref{cl3}) and (\ref{beta2}), is less reliable at large Bjorken $x$, $x \geq 0.01$, since the diffractive cross sections are quite small at those $x$.

We would like to stress that a  decrease of $l_{c}^{eff}$ with increasing $Q$ is a model-independent statement. It is a direct consequence of the factorization theorem in diffractive DIS and the QCD evolution of diffractive parton densities. As $Q$ increases, the parton densities become softer, i.e. they shift towards lower $\beta$ (or higher $M_{X}$ according to Eq.\ (\ref{beta})). Hence, at fixed $x$, $\langle \beta \rangle(Q^2,x)$ decreases as $Q$ is increased, or, alternatively, $\langle M_{X}^2 \rangle$ increases as $Q$ is increased. Hence, by virtue of Eq.\ (\ref{cl1}) $l_{c}^{eff}$ decreases with $Q$.

Such a behaviour of the effective coherence length is supported by the experiment. The ZEUS data on single-diffraction \cite{ZEUSsd} demonstrates that as $Q^2$ increases, diffractive states of large mass become increasingly important \cite{PW99}. Hence, according to Eq.\ (\ref{cl1}), the effective coherence length should decrease as  $Q^2$ increases due to increasing $\langle M_{X}^2 \rangle$ with $Q^2$.

In conclusion, we would like to point out that the present-day knowledge of diffractive quark and gluon parton densities enables one to evaluate  the leading twist contribution to nuclear shadowing in the 
quark and gluon channels separately.  In the 
nuclear rest frame, 
the quark channel corresponds to quark-antiquark fluctuations of 
the incoming virtual photon, 
while the gluon channel 
can be probed via a study of the
 quark-antiquark-gluon fluctuations of the incoming photon \cite{Abr}.
The fit of Ref.\ \cite{ACW}  suggests that the shapes of
the $\beta$-distributions of the diffractive gluon and quark densities at low $Q^2$, $Q^2 \sim 4$ GeV$^2$, are similar.
 This  leads to the similar 
coherence lengths $l_{c}^{eff}$
(up to the effects related to the region of small $\beta$, which are somewhat different in the two cases).
 With an increase of $Q^2$,  
larger scaling violations in the gluon channel 
lead to a faster decrease with $Q^2$ of $l_{c}^{eff}$  for the gluon channel than for the quark channel.
The detailed $Q^2$ dependence of $l_{c}^{eff}$
 for the gluon channel will be presented elsewhere.

\subsection{The dynamics of compact systems} 
\label{subsec:Dynamics}

As discussed in the 
Introduction,
at high energy, a projectile consists of coherent quark-gluons
 configurations of very different transverse spatial sizes. 
One can select rare processes where 
the small-size, or compact,   
quark-gluon fluctuations either dominate the process as in electroproduction of vector
 mesons by longitudinally polarized photons with high $Q^2$ or are  as important 
as the large-size configurations as in inclusive DIS or in diffractive processes initiated by
 transversely polarized photons. The cross section of the  interaction of the small-size
 configuration with the hadronic target is calculable in pQCD since small distances
 correspond to large 
$Q^2$,
where pQCD is applicable. It was found that the small-size configurations have reduced interactions with the target.

The cross section of the interaction of a compact $q\bar{q}$ fluctuation of the virtual photon with a hadronic target $T$ was derived 
in the leading $\alpha_{s}\, \log(Q^2/\Lambda^2)$ approximation in 
\cite{FKS96,FMS93,BBFS93,FRadS98}\footnote{This formula is also  valid in the leading $\alpha_{s}\,\log x$
approximation.} 
\begin{equation}
\sigma^{q \bar{q}}_{T}(b^2,x)
= \frac{\pi^2}{3} b^2 \left[ x
G_T(x, \lambda/b^2) \right] \alpha_{s}(\lambda/b^2) \ ,
\label{cs}
\end{equation}
where $b=(b_{q}-b_{\bar{q}})$ 
is the transverse inter-quark distance within the wave function of 
the photon; $\alpha_{s}(\lambda/b^2)$ is the QCD coupling constant; $G_{T}(x, \lambda/b^2)$ is the gluon parton density of the target.
This formula is a consequence of gauge invariance, asymptotic freedom, and the factorization theorem in QCD. The QCD evolution is properly taken into account through the gluon distribution of the target $G_{T}(x, Q_{eff}=\lambda/b^2)$. The scale  factor $\lambda$ can be estimated from the analysis of the 
$\sigma_{L}(\gamma^{\ast}N)$ cross section \cite{FKS96}. $\lambda$ slowly increases 
with the decrease of $x$ due to the  increase of the gluon density at small $x$.  For $x \sim 10^{-3}$ and $Q^2 \approx 10$ GeV$^2$,  $\lambda \approx 10$. For a recent detailed study of an approach to inclusive DIS based on Eq.\ (\ref{cs}), see \cite{FGMS99}.

For a colorless dipole consisting  of two gluons, the dipole cross section will be larger by a factor of $9/4$, which is  the ratio of the Casimir operators
of the SU(3)$_{c}$ color group for the octet and triplet representations \cite{rev2}.
Note that this observation is consistent with the analysis of DIS data in Ref.\ \cite{FS99} which demonstrated that the effective dipole cross section 
in  the gluon channel  is larger (by approximately a factor of 3) than in the quark channel.

The gluon density of the target in Eq.\ (\ref{cs}) is a feature that is absent in the two-gluon exchange models \cite{Low}. The two-gluon description in \cite{Low} is incomplete because at small Bjorken $x$, a compact quark-gluon configuration interacts with the target by producing 
 a soft self-interacting multi-gluon field.
The effect of this soft gluon field is expressed by the gluon parton density of the target $xG_{T}(x, \lambda/b^2)$.

The proportionality of 
 $\sigma^{q \bar{q}}_{T}(b^2,x)$ to the gluon density $  xG_{T}(x, \lambda/b^2)$
has found many experimental confirmations in inclusive DIS and in hard exclusive reactions  on nucleon and nuclear targets.   

Let us consider the  implications of Eq.\ (\ref{cs}) for inclusive DIS scattering and for the
 famous approximate Bjorken scaling for the total photoabsorption cross section 
$\sigma_{tot}(\gamma^{\ast}+p \to X)(Q^2,x)$. According to Eq.\ (\ref{cs}), 
the cross section for the interaction of a 
 compact quark-gluon configurations with a  target is  proportional to the
transverse size $b^2 \propto 1/Q^2$ of the compact configuration.
 It is important to note that this non-trivial result contradicts our experience from the  models, which existed before QCD. For example, in  the meson cloud  models of the  photon wave function,
all meson configurations  interact with the target with typical hadron-nucleon (nucleus) cross sections. 
In the case of DIS on nuclei at not extremely small $x$, 
such a behaviour leads to  $\sigma_{tot} \propto \ln x$ 
at $x \ll 1 / 2R_{A}m_{N}$ (see Eq.\ (\ref{cl1b})), which  
 contradicts the experimentally observed
Bjorken scaling and causes a paradox. The paradox is known as the Gribov paradox.

QCD explains the origin of the Gribov paradox.
Indeed, in pQCD, the contribution of small-size configurations to $\sigma_{tot}(\gamma^{\ast}+p \to X)(Q^2,x)$ behaves as $1/Q^2$ according to Eq.\ (\ref{cs}). As a result, it has the same $Q^2$ dependence as the
 non-perturbative contribution of  large-size configurations to $\sigma_{tot}(\gamma^{\ast}+p \to X)(Q^2,x)$, which   is suppressed by a factor proportional to $1/Q^2$ since the large-size  configurations occupy 
a small fraction
$\sim k_t^2/Q^2$
 of the phase volume. Therefore, the sum of these two contributions behaves approximately as $1/Q^2$   at large $Q^2$ in accordance with the approximate Bjorken scaling. 

The QCD prediction of Eq.\ (\ref{cs}) and its experimental confirmation both demonstrate that the existence of quark-gluon configurations, interacting weakly with hadron 
targets, is a fundamental property of QCD.

 Eq.\ (\ref{cs})  substantiates the pre-QCD  hypothesis of J.~Bjorken,  which became known as the aligned jet model  \cite{Bj}.   
 Within the framework of the parton model, Bjorken offered a solution  of  the Gribov paradox by suggesting  that the interaction of
 the quark and anti-quark 
with large relative transverse momenta $p_{t}$,
which form the photon wave function,  with  hadronic targets is suppressed by some  factor $f(\mu^2/(p_{t}^2+\mu^2))$, where $\mu$ is a cut-off scale. 
Since large relative momenta correspond to small relative distances, the aligned jet model foresaw 
the suppression of the interaction of the spatially small quark-antiquark dipole.
So, in the aligned jet model, only large-size configurations take part in the interaction with the target. Since the available phase space of the aligned jet configurations is proportional to $1/Q^2$, the scaling of $\sigma_{tot}(\gamma^{\ast}+p \to X)(Q^2,x)$ is restored in the aligned jet model. 

In QCD, Bjorken's  aligned jet model needs to be modified.
Firstly, Bjorken scaling of $\sigma_{tot}(\gamma^{\ast}+p \to X)(Q^2,x)$ is only approximate due to the QCD evolution with $\log(Q^2)$ of parton densities of the target.
 Secondly, as one can see from Eq.\ (\ref{cs}), the interaction of small configurations is suppressed but  not negligible. Thirdly, 
as a result of the QCD evolution, the gluon parton density of the target becomes large at small $x$, which makes the cross section sizable even for relatively small $b$. Fourthly, one has  to account for 
the suppression of on-mass-shell quark production and for the
compensation of this suppression by the radiation 
of hard gluons ( see, e.g.,  \cite{FS1,DKMT}).

Leading twist nuclear shadowing  predicted by the QCD-improved 
aligned jet model has been observed in inclusive DIS scattering on light and heavy nuclei, see  \cite{CERN1,CERN2}  and references therein.

\subsection {The factorization theorem for hard exclusive meson electroproduction off hadronic targets}
\label{subsec:FE}

A crucial element of perturbative calculations in QCD are  factorization theorems. Such theorems  prove that  amplitudes of certain hard processes can be 
computed as the convolution of the hard part, which can be computed in pQCD, and of the soft, non-perturbative, part, which contains information about parton distributions in the target.

Recently J.~Collins {\it et al.} \cite{Collins} have proved  the 
QCD factorization theorem for hard exclusive meson production in the reaction 
$\gamma^{\ast}_{L}+T \to M +T^{\prime}$, where $M$ is a meson 
($\rho$, $\pi$, $K$, $\eta$, $\dots$)
and $T$ and $T^{\prime}$ are baryon states of a fixed mass. This theorem is valid for 
hard processes initiated  by a highly virtual longitudinally polarized 
photon on any hadron target at fixed $x$ and 
$Q^2 \gg \Lambda_{QCD}^2$. The amplitudes of such processes are calculable as 
the convolution
of the hard kernel $H_{ij}(Q^2x_{1}/x,Q^2,z,\mu^2)$,
calculable in pQCD for a given  minimal Fock component 
of the light-cone wave function of the
meson wave function $\phi_{j}^{M}(z,\mu^2)$, and the skewed parton densities of the target $f_{i/p}(x_{1},x_{1}-x,t,\mu^2)$ (see Fig.\ \ref{fig:panic} for the schematic representation of the amplitude for the process $\gamma^{\ast}_{L}+T \to M +T^{\prime}$)
\begin{equation}
A=\sum_{i,j} \int^{1}_{0} dz  \int dx_{1} f_{i/p}(x_{1},x_{1}-x,t,\mu^2)\, H_{ij}(Q^2x_{1}/x,Q^2,z,\mu^2)\, \phi_{j}^{M}(z,\mu^2) \ .
\label{excl1}
\end{equation}
In Eq.\ (\ref{excl1}), $z$ is the fraction of the photon longitudinal momentum, carried by the quark or antiquark that constitutes the meson $M$; $\mu^2$ is the factorization scale which is usually set to $Q^2$;  $x_{1}$ and $x_{1}-x$ are the longitudinal light-cone momentum fractions of the partons connecting the hard and soft parts of the amplitude $A$. Note that the
skewed parton densities evolve as a function of $x_1$ and  $Q^2$  for fixed  $x=Q^2 / 2 (p \cdot q)$, while
the conventional parton densities evolve as a function of $x$ and  $Q^2$. In order to emphasize this difference, we keep a different 
notation for the
 skewedness parameter $\Delta \equiv x_{1}-(x_{1}-x)=x$.

A similar QCD factorization theorem can be derived  if $M$ and $T$ are baryons and 
$T^{\prime}$ is a meson.

Applications of the factorization theorem 
give a practical possibility of  calculating  cross sections of
hard exclusive processes in the same sense as total cross sections of inclusive DIS \cite{Collins,Radyushkin96,FKS96}. 
The kernels for the  QCD evolution of the
skewed (nondiagonal, nonforward)
parton densities
have been calculated in the leading $\alpha_{s}\, \log(Q^2/\Lambda^2)$ approximation
in  \cite{Braun,FG} and in the next-to-leading $\alpha_{s}\, \log(Q^2/\Lambda^2)$
approximation in  \cite{Belitsky,BFM99}. Thus, the major difference between  computing cross sections of hard diffractive processes and  more familiar inclusive  DIS processes
consists in the necessity to use the skewed parton densities instead of the
conventional parton densities.

 A natural assumption for the input skewed  parton densities $f_{i/p}(x_{1},x_{1}-x,t,\mu^2)$ at the initial scale $\mu=Q_{0}$  of the QCD evolution and small $x$ is  that they  should be taken equal to  the usual parton densities at $Q_{0}$ \cite{FG}. This conjecture is based on the observation that the effect of skewedness (nondiagonality) should be negligible at large $x_1$ and small $Q_{0}$ from which the QCD evolution starts, i.e. one can neglect the effect of $\Delta$
 at the input scale $Q_{0}$. 
 We refer the reader to Subsec.\ \ref{subsec:trajectory} where this question is investigated in detail by analyzing general tendencies of the QCD evolution in the $x-Q$-plane.

Practical ways to estimate skewed parton densities through conventional parton densities were suggested in Ref.\ \cite{FG,Rad2}. In particular,  using symmetries of the kernels of the skewed QCD evolution, it was proposed  in Ref.\ \cite{rev2,Rad2}
 that the skewed parton densities at given $x_{1}$ and $x$ are expressed through the conventional parton densities computed at  smaller $x^{\prime}=x_{1}-\Delta/2$.
The numerical analysis of \cite{FG2} showed that, indeed, it is more appropriate to compare the skewed parton densities to the usual parton densities taken at $x^{\prime} < x_{1}$.

The factorization theorem for hard exclusive meson electroproduction
off nucleon or nuclear targets enables one to experimentally measure 
the skewed parton distributions in 
the nucleon (nucleus) and the distribution of bare quarks within 
  mesons and nucleons (the minimal Fock components in the wave functions
 of mesons and nucleons). The number of  the skewed
parton distributions is much larger than the number of 
parton distributions measured in inclusive DIS since the former includes, for example, 
transitions like $N \to \Lambda$, various distributions of spin, etc.
Thus, it is reasonable to expect  significant progress in
the understanding of the structure of baryons and mesons from the investigation of hard diffractive processes.

\section{Hard diffractive processes}
\label{sec:HD}

In this section, we discuss the implications of 
color coherence for hard diffractive processes on nucleons and nuclei.

\subsection {The phenomenon of Generalized Color Transparency}
\label{subsec:GCTP}

The phenomenon of the  complete disappearance of initial and 
final state interactions  of spatially small-size wave packets
of quarks and gluons, produced in hard exclusive processes,
is usually referred to as the phenomenon of color transparency. 
Initial theoretical ideas on the color transparency phenomenon were intuitive and 
based on model assumptions. For the history of the development of this subject 
and appropriate references, see, e.g.,\ \cite{rev1}. The recent progress in 
the theoretical understanding of the color transparency phenomenon 
is based on the systematic application of the QCD factorization theorem to 
 hard exclusive and semi-exclusive reactions. 
In particular, the application of the factorization 
theorem to hard diffractive vector meson $V$  production on nuclear and nucleon targets at sufficiently small $x$ enables one to express the ratio of the forward differential cross sections for the nucleus and the nucleon  as 
(compare to Eq.\ (\ref{excl1}))

\begin{eqnarray}
&&\frac{d\sigma^{\gamma^{\ast}+A \to V+A}/dt(t=0)}
{d\sigma^{\gamma^{\ast}+N \to V+N}/dt(t=0)} = \nonumber\\
&&\frac{\sum_{i,j} \int^{1}_{0} dz  \int dx_{1} f_{i/A}(x_{1},x_{1}-x,t=0,Q^2)\, H_{ij}(Q^2x_{1}/x,Q^2,z)\, \phi_{j}^{V}(z,Q^2)}{\sum_{i,j} \int^{1}_{0} dz  \int dx_{1} f_{i/N}(x_{1},x_{1}-x,t=0,Q^2)\, H_{ij}(Q^2x_{1}/x,Q^2,z)\, \phi_{j}^{V}(z,Q^2)} \ ,
\label{excl2}
\end{eqnarray}
where $f_{i/A}(x_{1},x_{1}-x,t,Q^2)$ is the nuclear skewed parton density of parton $i$.

The phenomenon, described by Eq.\ (\ref{excl2}), can be called  generalized  color transparency \cite{talk1,talk2,Brod94}. Although it resembles  color transparency, the two phenomena 
lead to qualitatively different patterns of $A$ dependence for sufficiently small $x$.
Indeed, in case of the color transparency phenomenon, the complete disappearance of the initial and final state interactions
of the $q\bar q$ pair, when $x$ decreases,
 is predicted \cite{Bertsch}. On the contrary,
 in the  case of  the  generalized  color transparency  phenomenon, 
 the initial  state interactions  with nuclear medium do not disappear
 but increase with a decrease of $x$. 
 They are included into  nuclear modifications of 
 parton distributions (nuclear shadowing of parton distributions) 
 the nuclear target $f_{i/A}(x_{1},x_{1}-x,t,Q^2)$  (see, e.g., Fig. 
\ \ref{fig:shadrhobl}).

 It is important to repeat that the considered example of diffractive production of vector mesons 
  presents  a particular case of the 
QCD factorization theorem which states that nuclear effects in  
hard exclusive processes at  small $x$ are calculable in QCD 
within the framework of the generalized (skewed) QCD evolution equation 
for the parton distributions of nuclei.

Now let us consider two limiting cases of Eq.\ (\ref{excl2}).
  In the limit of fixed $x$ and $Q^2 \to \infty$, the
nuclear effects in the parton distributions disappear, i.e. $f_{i/A}(x_{1},x_{1}-x,t,Q^2) \to A f_{i/N}(x_{1},x_{1}-x,t,Q^2)$. This leads to the more familiar phenomenon of color transparency 
\begin{equation}
\frac{d\sigma^{\gamma^{\ast}+A \to V+A}/dt(t=0)}
{d\sigma^{\gamma^{\ast}+N \to V+N}/dt(t=0)} = A^2 \ .
\label{excl3}
\end{equation}

Eq.\ (\ref{excl3}) is a consequence of the disappearance of nuclear effects in the skewed  nuclear parton densities as a result of the  QCD evolution. The reason for such a disappearance is that the skewed (as well as conventional) parton densities at given $x$ and $Q^2$ are determined by  the input parton distributions at significantly larger $x$ and smaller $Q_{0}^2$ where nuclear shadowing is unimportant. Thus, as one evolves from $Q_{0}^2$ to $Q^2$, the unshadowed large-$x$ region contributes at smaller $x$ and, consequently, diminishes the effect of shadowing.  It explains how color transparency takes over generalized color transparency
in the limit of fixed $x$ and $Q^2 \to \infty$. We also refer the reader to Subsec.\ \ref{subsec:trajectory}, where general trends of the QCD evolution in the $x-Q$-plane are discussed on the quantitative level.

In order to illustrate the onset of color transparency numerically, we perform
an analysis of the QCD evolution of the skewed and conventional nuclear parton densities. Details of the analysis are presented in Subsect.\ \ref{subsec:NA}.
The limiting case, when $x$ is fixed and $Q^2 \to \infty$, is presented in Figs.\ \ref{fig:ratq} and \ref{fig:ratnd}.
One can see that, 
 at $x=10^{-3}$ and large $Q^2$, the ratio of nuclear to nucleon skewed and conventional gluon parton densities increases and will eventually
approach  unity. This observation  implies that the effect of nuclear shadowing vanishes and Eq.\ (\ref{excl3}) is justified in the discussed limit.

In the other limit of fixed $Q^2$ and $x \to 0$, QCD predicts 
that nuclear effects begin to play an important role. This occurs due to the 
growing importance of nuclear shadowing at small $x$. This effect can also be
interpreted as
the decrease of the influence of color transparency.
  This  limiting case  for the conventional  gluon parton densities is presented in Fig.\ \ref{fig:ratx}. The ratio of the  nuclear to nucleon conventional  gluon parton densities becomes smaller at fixed $Q^2$ and $x \to 0$, which 
demonstrates the increase of nuclear shadowing.

Within the models, inspired by the two-gluon exchange model \cite{Low}, i.e.  
by the analysis of the lowest-order diagrams of pQCD 
(see Ref.\ \cite{Bertsch}, for example),
 shadowing 
for the interaction of small dipoles is a higher twist effect which
decreases  $\propto 1/Q^2$
 when $Q^2$ increases. The difference between the predictions of \cite{Bertsch}
 and our analysis is due to 
the lack of Gribov diffusion to larger size configurations
in the parton ladder   in the lowest-order Feynman diagrams \cite{Gribov2,Gribov3}. Gribov diffusion has been observed long ago as a shrinkage of the
diffractive peak in $p\,p$ collisions with increasing  energy. Another cause for the difference 
is that the QCD evolution
equation predicts increasing parton densities at small $x$. This feature
is also absent
 in the lowest-order Feynman diagrams.

\subsection{A brief review of experimental data}
\label{subsec:BER}

The current subsection is concerned with hard exclusive reactions in the kinematic limit  where color transparency plays a dominant role. Namely, these reactions are vector meson electroproduction off  
nuclear targets, coherent photoproduction of $J/\psi$,
pion diffractive dijet production on nuclei and processes where the two-gluon exchange is prohibited. 

Exclusive meson production by longitudinal photons was  discussed in the previous subsection. Eq.\ (\ref{excl2}) represents the ratio of the corresponding forward differential cross sections.
In the limit of fixed small 
$x$ and $Q^2\to \infty$, Eq.\ (\ref{excl2})  has been proven in QCD  within the leading 
$\alpha_{s}\, \log(Q^2/\Lambda^2)\, \log x$ approximation in \cite{Brod94}  and within  the leading $\alpha_{s}\, \log(Q^2/ \Lambda^2)$ 
approximations in \cite{FKS96}. 

        Originally,  in \cite{FKS96,Brod94}, Eq.\ (\ref {excl2}) was obtained using the conventional gluon parton densities. The factorization theorem for  exclusive  vector meson electroproduction  requires that the skewed parton densities are used instead of the conventional parton densities. However, regardless  of whether  skewed or conventional parton densities of the target are used, Eq.\ (\ref{excl2}) leads to Eq.\ (\ref{excl3}) because 
in the  limit of fixed small 
$x$ and $Q^2\to \infty$ nuclear shadowing disappears in both  skewed and conventional parton densities, as we explained in Subsec. \ref{subsec:GCTP}.

An additional $A$ dependence of the cross section for  exclusive  vector meson electroproduction
 is also present.
 The cross section is proportional to the square of the nuclear form 
factor $F_{A}^2(t_{min})=\exp(-m_{N}^2 x_{Bj}^2 R_{A}^2/3)$, where $-t_{min}=m_{N}^2 x_{Bj}^2$ is the minimal momentum transfer to the target. However, due to 
the smallness of $x_{Bj}$ considered here, $x_{Bj} \ll \sqrt{3} / (R_Am_N) \approx 0.34 / A^{1/3}$, the effect of the nuclear form factor  is 
negligible in  Eqs.\ (\ref{excl2}) and (\ref{excl3}).

The behaviour,  predicted by Eq.\ (\ref{excl2}), has been verified experimentally
in a number of phenomena.
The E665 data  on exclusive $\rho$ meson electroproduction
off nuclear targets \cite{E665} reveals that the $A$ dependence of the cross section 
increases with the increase of $Q^2$.  Such a  behaviour  is a characteristic feature of the color transparency phenomenon. 
However, this result should  be considered with a  grain of salt  because  part of 
the $Q^2$ dependence can be attributed to the fact that in the
kinematics, covered by the E665 data, the coherence length $l_{c}$ is comparable
with the nuclear radius and that within the data sample $l_{c}$ depends on $Q^2$.

 Another example of processes, where color coherence plays a principal role, is coherent photoproduction of $J/\psi$.  The data on the cross section of coherent
photoproduction of $J/\psi$ off nuclei \cite{Sokoloff} at $\left<E_{\gamma}\right>=120$ GeV and $x_{eff} \approx 0.04 $,  where 
the shadowing effects in the gluon density are negligible, demonstrates that the cross 
section is proportional to $A^{n}$ with $n \approx {1.4}\pm{0.06}\pm{0.04}$.
This result is consistent with the idea in \cite{FS85} that $J/\psi$, produced 
in exclusive photoproduction processes, originates from conversion of
the photon to a
$c \bar c$ quark pair  at small relative distances, which hence interacts
with a small 
interaction cross section. At the same time, the  genuine $J/\psi$-nucleon cross section could be significantly larger  because, in case of $J/\psi$-nucleon interaction,  the $c \bar c$ quarks are at average distances within $J/ \psi$ \cite{FS85,Gerland}. 

Hard diffractive electroproduction of vector mesons, described above, is dominated by small-size quark-gluon configurations, 
enforced by choosing the special kinematic condition of $Q^2 \to \infty$. 
This condition selects small transverse size $q\bar{q}$ configurations in longitudinal photons.
Another way to select small  transverse size configurations,
 is to choose a special final state. An example of such a process  is coherent diffractive production of mini-jets by pions in the reaction $\pi+A \to jet(k_{t})+jet(-k_{t})+A$. If we impose the restriction that only high transverse momentum $k_{t}$ jets are produced and 
that only the diffractive peak is investigated such that the nucleus $A$ is intact,
 then it would insure only small $q\bar{q}$ configurations in the wave function of the incoming pions. 

The recent Fermilab experiment E791 data on diffractive 
dijet production by a pion projectile with momentum of 500 GeV/c
off nuclear targets of carbon and platinum  \cite{E791} is consistent with the predictions 
of  the generalized color transparency \cite{FMS93}.
  The analysis of  the E791 data demonstrates that  
$\sigma(\pi +A \to 2jets +A) 
\propto
A^{1.61 \pm 0.08}$ for $1.5 
\mbox{ GeV/c }
< k_{t} < 2.0$ GeV/c and  $\sigma 
\propto
A^{1.65 \pm 0.09}$ for 
$2.0 
\mbox{ GeV/c }
< k_{t}$ 
\cite{Ashery}. These results are consistent  with the predictions based on color transparency.

It is important to stress  that this $A$ dependence is
significantly faster 
than  the expected $A$ dependence for usual soft diffraction discussed in 
Sect.\ \ref{sec:SD}. Namely, the ratio of the  cross sections for platinum and carbon, $\sigma_{Pt} / \sigma_{C}$
 observed in the E791  experiment, is approximately seven times larger than this ratio, estimated within the framework of soft diffraction, where $\sigma \propto A^{2/3}$. 

In addition to the $A$ dependence, the E791 experiment confirmed other expectations for the color transparency regime: i) The pion wave function at $Q^2 \approx$ 10 GeV$^2$ ($k_{t} > 1.5$ GeV/c) is described well by the asymptotic wave function calculable in pQCD, $|\Psi_{\pi}(z)|^2 \propto z^2(1-z)^2$, where $z$ is the fraction of the longitudinal momentum of the pion carried by a quark in the infinite momentum frame \cite{Efremov80,Lepage80}; ii) The $k_{t}$ dependence of the cross section at $k_{t} > 1.5$ GeV/c 
is also consistent with the pQCD calculation of \cite{FMS93}.

It is important to note that the observed $A$ dependence of the cross section of diffractive dijet production  is close to that found for 
 exclusive photoproduction of $J/ \psi$ \cite{Sokoloff}. This observation is 
in line with the QCD factorization theorem of \cite{Collins} since
 the  cross sections for  diffractive dijet production and exclusive photoproduction of $J/ \psi$ are expressed through the same nuclear skewed
gluon distribution  at similar light-cone gluon momenta. 

        More experiments are needed to understand the dynamics of hard diffractive exclusive processes more clearly. We suggest  measuring  diffractive dijet production off nuclei 
 by different projectiles, such as kaons, or to measure  charmed jet production by real photons.

Eqs.\ (\ref{excl2}) and (\ref{excl3}) 
are very general. They have been  proven using the QCD factorization theorem and the QCD evolution equation. According to the  QCD factorization theorem for certain hard processes 
dominated by the scattering of compact system
from
hadron (nuclear) targets,
the cross sections are expressed through the target parton (gluon) densities.
But, as it follows from the QCD evolution equation,  nuclear shadowing
vanishes at fixed $x$ with the increase of $Q^2$. Hence, in this limit,  Eq.\ (\ref{excl3}) is justified. 
A similar prediction
 is also applicable to the processes where 
the exchange with vacuum quantum numbers in $t$ channel is forbidden and where
as a result 
the two-gluon exchange does not contribute.
Below we will give a few examples of such processes.

 The QCD factorization theorem predicts
the color transparency phenomenon for the zero-angle charge-exchange coherent processes, initiated by longitudinal photons
in the limit of $Q^2\to \infty$ and $1 /(2 m_Nx) \gg R_A$
(so that the effect of $t_{min}$ in the nuclear form factor is small)
 but $l_{c}^{eff} \ll r_{NN}$, where $r_{NN}$ is the average
 internucleon distance in the nucleus
\cite{talk1}  

\begin{equation}
\frac{d\sigma(\gamma_{L}^{\ast}+A\to \rho^{-}+A')/dt}
{d\sigma(\gamma_{L}^{\ast}+B\to \rho^{-}+B')/dt}=
 \frac{|\langle A|T_{+}|A^{\prime} \rangle|^2}{|\langle B|T_{+}|B^{\prime} \rangle|^{2}} \ ,
\label{nr1} 
\end{equation}
where $A^{\prime}$ ($B^{\prime}$) is the isobaric state of a nucleus with the atomic number
$A$ ($B$) with a neutron  substituted by a proton; $T_{+}$ is the isospin-raising operator.

In order to diminish the sensitivity of the discussed processes to the nuclear wave function of the isobaric states, it can be
compared with the 
 zero-angle charge-exchange coherent processes initiated by 
high-energy
neutrinos or anti-neutrinos:
\begin{equation}
\frac{d\sigma(\nu_{l}+A\to l+A')/dt}
{d\sigma(\nu_{l}+B \to {\l} +B')/dt}=
\frac{|\langle A|T_{+}|A^{\prime} \rangle|^2}{|\langle B|T_{+}|B^{\prime} \rangle|^{2}} \ ,
\label{nr11} 
\end{equation}
where $\nu_{l}$ is the neutrino corresponding to the lepton $l$.

Another example of  zero-angle charge-exchange coherent processes, where color transparency dominates, is the production of 
electrically charged pseudo-scalar mesons \cite{talk1}. 
In the limit of $Q^2 \to \infty$ and $l_{c}\geq R_A$ but $l_{c}^{eff} \ll r_{NN}$, 
QCD predicts for the reactions, initiated by virtual photons
\begin{equation}
\frac{d\sigma(\gamma_{L}^{\ast}+A\to \pi^{-}+A')/dt}
{d\sigma(\gamma_{L}^{\ast}+B\to \pi^{-}+B')/dt}=
\frac{|\langle A|T_{+}|A^{\prime} \rangle|^2}{|\langle B|T_{+}|B^{\prime} \rangle|^{2}} \ .
\label{nr2} 
\end{equation}

 It is important to emphasize that in the processes, described by Eqs.\ (\ref{nr1}), (\ref{nr11}), and (\ref{nr2}),
 exchange by quarks is required.

Similar predictions are applicable for the 
coherent production of $\pi^{+}$, $\rho^{+}$ and $f$ mesons and pairs of mesons.

The QCD factorization theorem \cite{Collins}  can be generalized to semi-inclusive 
processes off nuclei with  large rapidity gap between the
photon and target fragmentation regions. For zero-angle scattering, at $Q^2 \to \infty$
and fixed  $x$,
 the onset of  color transparency is expected 
\begin{equation}
{d\over dt}\sigma(\gamma_{L}^*+A\to M+Rap. \, Gap+X')=A
{d\over dt}\sigma(\gamma_{L}^*+N\to M+Rap. \, Gap+X') \ .
\label{nr3} 
\end{equation}

  It is necessary to stress that in practical applications taking account of the leading
twist effects in Eqs.\ (\ref{excl2}), (\ref{excl3}), and (\ref{nr3}) is insufficient since the small transverse size wave packet produced
tends to expand to a normal hadronic size. In order 
to suppress this effect, one needs to impose the specific kinematic conditions --  rather large $Q^2$ or sufficiently
small $x$ -- so that the expansion will happen outside of the nucleus.

\subsection{Hard diffractive processes at HERA}
\label{subsec:HDPH}
 
In this subsection, experimental results, obtained at HERA, are reviewed from the point of view of color coherence.

Hard diffractive electroproduction of vector mesons on hydrogen 
$\gamma^{\ast} +p \to V +p$, where $V=\rho$, $\phi$, or $J/\Psi$,
has recently been observed  at HERA \cite{HERA}. The
data  shows several phenomena, expected in  QCD. 
Firstly, QCD predicts the
increase of the cross section of hard diffractive 
processes, initiated by a projectile in a spatially small 
configuration of the size $1/Q^2$, as a
power of the invariant energy of the collision\footnote{Of course, one should use skewed parton densities in Eq.\ (\ref{hera1}). However, in the kinematics of the experiments,  the numerical difference between the skewed and conventional parton densities is rather small.}
\begin{equation}
\sigma(\mbox{hard diffractive})\approx \Big(xG_p(x,Q^2)\Big)^2 \ .
\label{hera1} 
\end{equation}
Such  a behaviour has been proven within the leading $\alpha_{s}\,\log x$
approximation in \cite {Brod94,FS89} and within the leading 
$\alpha_{s}\,ln Q^2/\Lambda^2$ approximation in \cite{Collins,FKS96,FMS93}.
The result of Eq.\ (\ref{hera1}) is in  broad agreement with
the  energy dependence of cross sections for  diffractive
electroproduction of vector mesons $\rho$, $\phi$, and $J/\psi$ 
observed at HERA \cite{HERA}. The rapid increase of the cross 
sections of hard diffractive processes with increasing energy is one of the signatures
of the color transparency  phenomenon \cite{Brod94}.

Secondly, the data have confirmed other QCD predictions for  hard exclusive diffractive electroproduction of 
vector mesons such as   the enhancement of heavy flavor vector 
mesons
production (in contrast to moderate $Q^2$ processes, in which
production of heavy flavors is suppressed), dominance of the 
longitudinal over transverse  cross sections at large $Q^2$, the universal dependence 
of cross sections of hard diffractive processes on the momentum 
transfer $t$, large cross sections of diffractive production of
excited states of vector mesons, restoration of the  SU(3) symmetry prediction  for the ratio of cross sections for different vector mesons in the final state,  etc. For the review and references
see \cite{rev1,rev2,HERA}. The current data is  consistent with the expectation  of
the onset of the hard QCD regime for hard diffractive processes 
at large $Q^2$ \cite{HERA}. Hence, the discovery of a variety 
of new calculable in QCD hard phenomena  is expected. 

 An important remark here is in order. All the 
above discussed
hard exclusive diffractive processes
were initiated by longitudinal photons. In such processes, the hardness of the photons (when the photons have large $Q^2$) leads to color transparency. This logic is inapplicable to the processes, dominated by transversely polarized photons. In case of transverse photons, the soft component of the wave function (large size) may be as important as the hard component 
(small size).
However, as a result of the Sudakov-type form factors, contributions 
of the soft component are suppressed at very large $Q^2$.
 The relative importance of small and large dipole sizes in exclusive diffraction, initiated by transversely polarized photons, requires further studies.

The importance of the soft component of the transversely polarized  virtual photons
 manifests itself in   
the discovery of significant cross sections
of diffraction in DIS \cite{HERA} and in the leading twist nuclear 
shadowing in DIS \cite{CERN1,CERN2}. For these processes, the hardness  does not necessarily mean color transparency.
The recent data of HERA \cite{HERA} suggests that, for $Q^2 \geq 6$ GeV$^2$,
exclusive electroproduction of vector mesons is dominated by
the compact configuration of the photon wave function. If so,  generalized color
transparency will be applicable to coherent diffractive
electroproduction of transversely polarized vector mesons at these $Q^2$.

\subsection{Hard diffractive electroproduction of photons}
\label{subsec:DVCS}

The generalized vector dominance model (GVMD) is often considered as an alternative
to the QCD explanation of the inclusive and diffractive
DIS, for a recent review and references see \cite{Schil}.
  In the GVMD model, using the basis of hadronic states,
the 
(virtual) photon interaction with the nucleon proceeds via the transition of the photon  into different vector meson states, when the diagonal $V+N \to V+N$ and non-diagonal 
$V+N\to V^{\prime}+N$ transitions are accounted for.

It was argued in \cite{FGS} that the GVMD models are successful in the theoretical description of photoproduction and electroproduction at high energies since they
 possess (in an approximate way) basic features of theories with color coherence. 

However, it turns out that 
the process of diffractive electroproduction of photons in the DIS  reaction 
$\gamma^{\ast}(Q^2)+p \to \gamma^{\ast}(Q_{f}^2)+p^{\prime}$, which
has been a subject of recent  theoretical and 
experimental studies, can discriminate between the GVMD and pQCD pictures of DIS.
 Within the framework of pQCD, the ratio of the imaginary parts of the 
amplitudes of the real and virtual photon electroproduction  was found 
to be approximately independent of $Q^2$ at small $x$ \cite{FFS}
\begin{equation}
\frac{{\rm Im} A(\gamma^{\ast}(Q^2)+p \to \gamma(Q^2=0)+p^{\prime})}
{{\rm Im} A(\gamma^{\ast}(Q^2)+p \to \gamma^{\ast}(Q^2)+p)} \approx 2 \ .
\label{extra2}
\end{equation}
The recent HERA data \cite{dvcshera} seems to confirm the result of Eq.\ (\ref{extra2}). 

Within the framework of the GVMD model, which contains non-diagonal transition between the neighbouring  vector mesons \cite{gvmd}, it was found in \cite{FGS} that   
\begin{eqnarray}
&&\frac{{\rm Im}A(\gamma^{\ast}(Q^2)+p \to \gamma(Q^2=0)+p^{\prime})}{{\rm Im}A(\gamma^{\ast}(Q^2)+p\to \gamma^{\ast}(Q^2))+p)}= \nonumber\\
&&\Big(\sum_{nm} \frac{e}{f_{n}}\frac{M_{n}^2}{M_{n}^2+Q^2}\Sigma_{nm}\frac{e}{f_{m}}\Big) \Big/ \Big(\sum_{nm} \frac{e}{f_{n}}\frac{M_{n}^2}{M_{n}^2+Q^2}\Sigma_{nm}\frac{e}{f_{m}} \frac{M_{m}^2}{M_{m}^2+Q^2}\Big) \ .
\label{extra1} 
\end{eqnarray} 
The comparison of predictions of Eqs.\ (\ref{extra2}) and  (\ref{extra1}) is presented in Fig.\ \ref{fracdvcs}. One can readily see that the result of the GVMD model is consistent with the pQCD result only in a limited range of $Q^2$, $1 \leq Q^2 \leq 5 \div 6$ GeV$^2$. Hence, the forthcoming HERA data on diffractive electroproduction of photons in the DIS reaction $\gamma^{\ast}(Q^2)+p \to \gamma^{\ast}(Q_{f}^2)+p^{\prime}$ in a wide range of $Q^2$,
would allow to 
 establish the limits of applicability of the complementary
description of DIS, which employs the  GVMD models with elements 
of
color transparency built-in.

\section{Numerical analysis of the DGLAP QCD evolution}
\label{sec:NA}

In this section, the onset of color transparency and generalized color transparency 
is studied numerically 
by performing the QCD evolution of parton densities of  nuclei. Adopting a reasonable algorithm, we also analyze the paths in the $x-Q$-plane along which the QCD evolution proceeds. The knowledge of these paths enables one to substantiate the choice of the initial condition for the  QCD evolution of the skewed parton densities and to explain the $Q^2$ evolution of nuclear parton densities. Also, the limits of applicability of 
the  QCD evolution for nuclear parton densities are estimated.

\subsection{The QCD evolution paths}
\label{subsec:trajectory}

The integro-differential DGLAP  evolution equations, which govern the $Q^2$ and $x$ dependence of parton densities, read as follows
\begin{eqnarray}
&&\frac{dq^{NS}(x,Q^2)}{d ln Q^2}=\frac{\alpha_{s}(Q^2)}{2 \pi} \int^{1}_{x} \frac{dy}{y} q^{NS}(y,Q^2) P_{qq}(\frac{x}{y}) \ , \nonumber\\
&&\frac{d}{d ln Q^2} \pmatrix{q^{S}(x,Q^2)\cr
g(x,Q^2)}=\frac{\alpha_{s}(Q^2)}{2 \pi} \int^{1}_{x} \frac{dy}{y} \pmatrix{P_{qq}(\frac{x}{y})&P_{qg}(\frac{x}{y})\cr
P_{gq}(\frac{x}{y})&P_{gg}(\frac{x}{y})} \pmatrix{q^{S}(y,Q^2)\cr
g(y,Q^2)} \ .
\label{dglap}
\end{eqnarray}

In Eqs.\ (\ref{dglap}), $q^{NS}(x,Q^2)=q_{i}(x,Q^2)-q_{j}(x,Q^2)$, with any $i \neq j$,  
is the quark flavor non-singlet distribution;
$q^{S}(x,Q^2)=\sum_{i}(q_{i}(x,Q^2)+\bar{q_{i}}(x,Q^2)$ is the quark flavor singlet distribution; 
$g(x,Q^2)$ is the gluon distribution;
$\alpha_{s}(Q^2)$ is the running QCD coupling constant; $P_{qq}(x/y)$, $P_{qg}(x/y)$, $P_{gq}(x/y)$, and $P_{gg}(x/y)$
are the splitting functions calculable in pQCD.

Eqs.\ (\ref{dglap}) are often solved numerically. One example is the brute force method, when iterations in $Q^2$ are used. In particular, knowing the parton densities at $Q^2_{0}$, one can find the parton densities at larger $Q^2$ using Eq.\ (\ref{dglap}) as
\begin{equation}
q^{NS}(x,Q^2)=q^{NS}(x,Q_{0}^2)+\frac{Q^2-Q^2_{0}}{Q^2_{0}}
\frac{\alpha_{s}(Q_{0}^2)}{2 \pi} \int^{1}_{x} \frac{dy}{y} q^{NS}(y,Q_{0}^2) P_{qq}(\frac{x}{y}) \ . 
\label{dglap2}
\end{equation}
Similarly, Eq.\ (\ref{dglap}) can be solved for the quark flavor
 singlet and gluon parton densities.
 By choosing the difference between $Q^2$ and  $Q^2_{0}$ small enough and repeating the procedure of  Eq.\ (\ref{dglap2}) a desired number of times, each time using  the parton densities from the previous step in $Q^2$, one finds the solution of Eq.\ (\ref{dglap}).   
 From Eq.\ (\ref{dglap2}) one can see that the parton densities at the input  
scale $Q^2_{0}$ determine the the parton densities at a 
larger scale $Q^2$, $Q^2 > Q^2_{0}$.

The general trends of the QCD DGLAP $Q^2$ evolution are well known. As $Q$ increases, the parton densities shift toward lower $x$ due to the emission of soft partons. So, the QCD evolution proceeds along a path 
 in the $x-Q$-plane, which extends from low $Q$ and high $x$ towards large $Q$ and small $x$. The detailed knowledge of this path is very important. It enables one, for example, to estimate the influence of the input parton densities at the initial scale $Q_{0}$ on the result of the QCD evolution at higher $Q$ and judge which regions of $x$ at  $Q_{0}$ contribute to the parton densities at $Q > Q_{0}$.
Thus, our study of the general trends of the QCD evolution addresses the issues of the choice of the input distribution for the QCD evolution for the skewed parton densities and  the $Q^2$ and $x$ dependence of nuclear shadowing. It might also contribute to our understanding of the limits of applicability of the leading twist QCD evolution equation.

In order to investigate a path in the  $x-Q$-plane relevant for the  QCD evolution,  the following algorithm was adopted. 
\begin{itemize}
\item{Fix the final point of the QCD evolution at some large $Q$ and small $x$. We have studied three different final points at $Q$=5 GeV and $x$=10$^{-4}$, 10$^{-3}$, and 10$^{-2}$. By solving Eq.\ (\ref{dglap2}) numerically, we find the parton densities at these final points.}

\item{Begin increasing the lower limit of integration in Eq.\ (\ref{dglap2}) from $x$ to some $x^{\prime}$. It will decrease the integration interval over $y$ at the initial scale $Q_{0}$, which contributes to the parton densities at the final points -- the values of the parton densities at the final points will decrease. Choose $x^{\prime}$ such that the integration from $x^{\prime}$ to 1 in Eq.\ (\ref{dglap2}) gives
the parton densities at the final points, which are half the parton densities, when the integration takes place from $x$ to 1. Then, for each final point, i.e. for  each pair $(Q>Q_{0},x)$,  $x^{\prime}$ represents  the average $x$ at $Q_{0}$. Take those  $x^{\prime}$ as the initial point of the QCD path.}

\item{Now that the initial and final points of the path are known, one needs to connect them, i.e. one needs to find pairs $(Q_{0} < Q < 5\  {\rm GeV},x^{\prime \prime})$ through which the path passes. At each $Q_{0} < Q < 5$ GeV, $x^{\prime \prime}$ is found by requiring that, when the integration in  Eq.\ (\ref{dglap2}) is from $x^{\prime}$ to 1, the parton densities at $x^{\prime \prime}$ are half the parton densities at $x^{\prime \prime}$,  when the integration takes place from $x$ to 1.}    

\end{itemize}

The numerical solution of Eq.\ (\ref{dglap2}) was found using
the CTEQ QCD evolution package. In order to illustrate the role, played by the shape of the input parton densities, two different input parameterizations of Ref.\ \cite{CTEQ} were considered: CTEQ4LQ with the starting scale of the evolution $Q_{0}=0.7$ GeV and CTEQ4D with $Q_{0}$=1.6 GeV. The parameterization CTEQ4LQ provides ``valence-type'' gluon and valence quark  initial distributions localized around $x=0.2$; the input sea quark distribution is quite small. CTEQ4D gives slightly rising input parton distributions as $x$ decreases.

The results for the path in the $x-Q$-plane for the gluon, $u$ and $s$ quark parton densities
 are presented in Fig.\ \ref{fig:pathlq}. 
 CTEQ4LQ was used as an  input. Each line represents a path, which
begins at $Q_{0}$=1.6 GeV and certain $x^{\prime}$ and
 ends at $Q$=5 GeV and $x=10^{-4}$, or $10^{-3}$, or $10^{-2}$. Each curve passes through $1.6 \leq Q \leq 5$ GeV and $x^{\prime \prime}$, defined by the above algorithm. In  Fig.\ \ref{fig:pathlq}, the upper panel corresponds to the implementation of our algorithm for the gluon parton density; the middle panel is for the $u$ quarks; the lower one is for the $s$ quark parton density.

From Fig.\ \ref{fig:pathlq}  one can see that the QCD evolution for gluons and quarks is different. The gluon density evolves in both $Q$ and $x$; the $u$ quark evolves mostly in $Q$ and turns sharply to reach the final point only at the end; the trajectory for the $s$ quark  is a mixture of the trajectories for the gluons and $u$ quarks.

Note that the cross-over of the paths for the $u$ quark in Fig.\ \ref{fig:pathlq} is due to 
the following reason. At small $x$, $x=10^{-4} \div 10^{-3}$, the $u$ quark parton density is rising and large. Hence, its support could come only from $x \approx 0.2$, where the input parton densities are peaked. This explains why the  initial points of the corresponding paths are located at large $x$. On the other hand, the $u$ quark parton density has a local minimum at $x \approx 10^{-2}$ and, thus, the large support region in Eq.\ (\ref{dglap2}) is not required.  This leads to the initial point at $x \approx 0.06$ for the path, which ends at $Q$=5 GeV and $x=10^{-2}$.

 Fig.\ \ref{fig:pathlq}  presents a $\it quantitative$ approach to the path of the  QCD evolution. One could arrive at the same $\it qualitative$ conclusions  simply by examining the corresponding parton densities at different $Q$ as a function of $x$.

 In the sense of the proposed algorithm,  Fig.\ \ref{fig:pathlq}  answers the question: In Eq.\ (\ref{dglap2}), which regions of $y$  at the input scale $Q_0$ are important at higher $Q > Q_{0}$? For example, one can see from Fig.\ \ref{fig:pathlq}  that at $Q$=5 GeV and $x=10^{-4}$, or $x=10^{-3}$ , or  $x=10^{-2}$ half of the gluon parton density has come from $x \geq 0.18 \div 0.22$ at $Q_{0}$=0.7 GeV.
This observation can be applied  to an analysis of 
the QCD evolution of skewed parton densities in the proton  and conventional  parton densities in nuclei.

Let us assume that in the kinematics of a certain exclusive process, 
the skewedness parameter $\Delta=x_{Bj}$ is small. Then, its effect is negligible at large values of $x_{1}$. We refer the reader to Subsect.\ \ref{subsec:FE} for the definition of the kinematical variables. 
Since half of the contribution at small $x_1$ and large $Q$ has come from $x_1 \geq 0.2$ at the input scale $Q_{0}$, the influence of the effects of $\Delta$ will be decreased by at least 50\% at small  $x_1$ and large $Q$. Therefore, this analysis justifies our claim that it is a good approximation to neglect the skewedness parameter $\Delta$ in the input for the skewed QCD evolution.

Applying the result of  Fig.\ \ref{fig:pathlq} to the QCD evolution of conventional  nuclear parton densities,   one can readily explain the decrease of nuclear shadowing with the increase of  $Q$. Since the major contribution at large $Q$ and small $x$ comes from larger $x$ at the input scale $Q_0$ where the effects of nuclear shadowing are not so significant, one would expect the effect of nuclear shadowing to decrease as $Q$ increases.

Fig.\ \ref{fig:pathlq} presents average paths. One can go one step further and study  the influence of various regions of $y$ in Eq.\ (\ref{dglap2}) on the resulting parton densities in the final points and not only the average $y$ that we called $x^{\prime}$ and $x^{\prime \prime}$ . Such a  study will help to answer the questions about the input for the skewed QCD evolution equation and the $Q^2$ behaviour of nuclear shadowing in more detail.

To this end, let us introduce functions 
$f_{g}(x,z)$ for gluons, defined as
\begin{equation}
f_{g}(x,z)=\frac{xG(x,Q=5)}{xG(x,Q=5)_{z}} \ ,
\end{equation}
and similar functions for quarks and antiquarks. Here
 $xG(x,Q=5)$ is the gluon density at $Q=5$ GeV  and $x=10^{-4}$, or $10^{-3}$, or $10^{-2}$; $xG(x,Q=5)_{z}$ is the gluon density at the same $Q$ and $x$ but originated from Eq.\ (\ref{dglap2}), where the lower limit of integration is $z$. Thus, by definition $f_{g}(x,x)$=1 and $f_{g}(x,1)$=0. 

$f(x,z)$ for gluon, $u$ and $s$ quark parton densities as a function of $z$  are presented in Fig. \ref{fig:displq}. It
helps to quantify further the importance of various regions of integration over $y$ in Eq.\ (\ref{dglap}).
For example,
from the upper panel  of  Fig. \ref{fig:displq}, one can see that at $Q$=5 GeV and $x$=10$^{-4}$, 80\% of the gluon density has come from $y \geq 0.1$ at $Q_{0}$=0.7 GeV.

We have also investigated how the QCD evolution path and the functions $f_{g,u,s}(x,z)$ are affected by the choice of the input distribution for the QCD evolution equation.
 Figs.\ \ref{fig:path4d} and \ref{fig:disp4d} 
represent the QCD evolution paths and functions $f_{g,u,s}(x,z)$ for the case, when the parameterization CTEQ4D was used as the input.
 One can see that the fact that CTEQ4D input parton densities increase slightly as $x$ decreases, affects the QCD evolution paths. While the general trend remains unchanged, the initial points for the paths shift to lower $x$. 

In Fig.\ \ref{fig:disp4d}, one can see that at $Q$=5 GeV and $x$=10$^{-4}$, 70\% of the gluon density has come from $y \geq 10^{-3}$ at $Q_{0}$=1.6 GeV. Thus, if the skewedness parameter $\Delta$ is of order $10^{-4}$, its effect is negligible at $x_{1} \approx 10^{-3}$. Therefore, Fig.\ \ref{fig:disp4d} quantitatively demonstrates that $\Delta$ can be safely neglected in the input for the skewed QCD evolution.

The comparison of Figs.\ \ref{fig:pathlq} and \ref{fig:path4d} 
demonstrates that the proposed algorithm for finding the path, along which the QCD evolution for the gluon and $s$ quark parton densities
 proceeds, does not depend on the initial  scale $Q_{0}$. Namely, the corresponding curves in  Fig.\ \ref{fig:pathlq} at $Q$=1.6 GeV pass through almost exactly the same $x^{\prime}$ as in Fig.\ \ref{fig:path4d} at  $Q_{0}$=1.6 GeV. Thus, Figs.\ \ref{fig:pathlq} and \ref{fig:path4d} can be used to analyze the QCD evolution of the gluons and $s$ quarks  between any two scales.
A radically different input for the valence quarks for the two parameterizations, considered here, makes the corresponding evolution paths in the region $1.6 \leq Q \leq 5$ GeV very different, and, thus, the universality discussed above does not hold.

In conclusion of this subsection, we would like to stress that our studies of the typical paths of the QCD evolution in the $x-Q$-plane presented in Figs.\ \ref{fig:pathlq} and \ref{fig:path4d} demonstrate that the input for the skewed QCD evolution can be taken as the input for the QCD evolution of conventional  parton densities as long the input densities are constant or grow insignificantly as $x \to 0$ at the initial evolution scale $Q_{0}$. Such a behaviour at low $x$ is constrained by the slow rise of cross sections for soft hadron-hadron interactions as energy increases.

\subsection{Numerical analysis of nuclear shadowing} 
\label{subsec:NA}

In this subsection we will present our numerical studies  of the $Q^2$ and $A$ dependence of skewed and ordinary
parton densities in nuclei and nucleons using the QCD DGLAP evolution equation.
The  studies serve the purpose of estimating numerically the regime of the
 onset of color transparency for hard exclusive 
and inclusive processes.

 As   explained in Sec.\ \ref{sec:CC}, the  
kinematics of exclusive reactions dictates the use of skewed
parton densities in evaluating 
cross sections of hard exclusive processes. In the limit of small  and 
fixed $x$ and  large $Q^2$, the onset of color 
transparency manifests itself as the absence of nuclear 
shadowing in the ratio 
\begin{equation}
R_{G}(x_{1},\Delta,Q)=\frac{xg_{A}(x_1,x_2,Q^2)}{A xg_{N}(x_1,x_2,Q^2)} \to 1 \ .
\label{r1}
\end{equation}

As discussed in Subsect.\ \ref{subsec:BER}, an additional $A$ dependence in Eq.\ (\ref{r1}) due to the nuclear form factor is negligible at small $x_{Bj}$ . Hence, the contribution due to the nuclear form factor  was omitted in our numerical analysis\footnote{Note also that for semi-inclusive processes when a sum over
nuclear fragmentation processes is taken, there is no suppression due to the nuclear form factor.}.

The numerical analysis of the conventional (non-skewed) QCD evolution equation was performed with the next-to-leading order in $\alpha_{s}\,\log(Q^2/\Lambda^2)$ accuracy (NLO) using the CTEQ QCD evolution package. For the  nucleon parton densities,  we used the CTEQ4D parametrization with the starting evolution point  $Q_{0}$=1.6
GeV. The choice was motivated by the observation that the CTEQ4D   gives a reasonable
nonperturbative input:
the gluon parton density is  a rather flat function of Bjorken 
$x$ at small $x$ in accordance with physical intuition based on  the slow rise with energy of soft hadron-hadron cross sections.

In order to carry out the numerical analysis of the skewed QCD evolution equation,  we used a modified version 
of the CTEQ QCD evolution package \cite{FG3}. In Subsect.\ {\ref{subsec:trajectory}, it is  explained why the input for the skewed QCD evolution can be taken  as being equal to the input for the QCD evolution for conventional parton densities. 
 Note that the detailed investigation of   this issue  showed that the input distribution should be fairly flat  for this conjecture to be valid.
 Consequently, we started 
our QCD evolution with the input skewed parton densities $g(x_1,x_2,Q^2_{0})=x_1G(x_1,Q^2_{0})$ \cite{FG}, where $G(x_1,Q^2_{0})$ is the conventional  parton density.
The accuracy of our calculations is of the leading order in $\alpha_{s}\, \log(Q^2/\Lambda^2)$ (LO) because the skewed evolution kernels of the evolution package  are implemented with the same accuracy.
Hence, for  the input nucleon parton densities we took  the leading order in $\alpha_{s}\, \log(Q^2/\Lambda^2)$  CTEQ4L parameterization with the starting evolution point  $Q_{0}$=1.6 GeV \cite{CTEQ}.

In order to account for significant nuclear shadowing at  $10^{-5} \div 10^{-4} \leq x \leq 0.05$ for gluon and sea quark nuclear parton densities and slight enhancement at $0.05 \leq x \leq  0.2$ of the gluon  nuclear parton density, we used the parameterization of Ref.\ \cite{FS99}.
This parameterization  is based on the recent analysis of diffractive $e\,p$ data of HERA, which  revealed  that the amount of shadowing in the gluon channel is approximately 3 times larger than  was thought before.

Since gluons and valence quarks evolve separately at small $x$, one does not need to model shadowing and antishadowing for the valence quarks when studying the gluon parton density. Should one need to study nuclear modifications of valence quarks, one can assume  that the valence quark parton densities are shadowed at small $x$ and then enhanced at larger $x$. However, one has to satisfy the baryon and momentum sum rules for the nuclear parton densities.

 Results of the QCD evolution for nuclei with $A=$12, 40, 100 and 200 are presented in Figs.\ \ref{fig:ratq}-\ref{fig:ratx}. Figs.\  \ref{fig:ratq} and \ref{fig:ratx} are the results of the QCD  evolution of the conventional  (diagonal) parton densities. The conventional QCD evolution is relevant for hard inclusive 
processes off nuclei such as DIS.

 Fig.\ \ref{fig:ratq} illustrates decreasing nuclear shadowing due to the QCD evolution at high $Q^2$ and fixed $x$.   
The ratio of the nuclear to nucleon gluon parton density per nucleon
$R_{G}(x,Q)=G_{A}(x,Q)/AG_{N}(x,Q)$ is plotted as a function of $Q$ at 
$x=10^{-3}$.
 One can see that the amount of  shadowing  for gluons,  given by 
$1-R_{G}(x,Q)$, decreases by a factor of 5 as $Q$ increases from 1.6 to 15 GeV for nuclei with $A=12$ and $A=40$ and by approximately a factor of $4.5$ for nuclei  with $A=100$ and $A=200$.

The onset of color transparency for exclusive processes (see also Eqs.\ (\ref{excl3}) and ({\ref{r1})) is presented in Fig.\ \ref{fig:ratnd}.
The ratio of nuclear to 
nucleon skewed gluon parton densities  $R_{G}(x_1,\Delta,Q)=g_{A}(x_1,x_2,Q)/Ag_{N}(x_1,x_2,Q)$ is plotted
as a function of $Q$ at fixed $x_1=10^{-3}$ for different values 
of $x_{1}-x_{2}=x_{Bj}=\Delta$.
One can see that the  curves
for different $\Delta$ for each $A$
 are remarkably close to each other.
This is an interesting result because, even though the skewed
gluon parton density is significantly larger than the diagonal one at small $x$, the effect of the 
skewedness $\Delta$
(asymmetry) cancels almost exactly in $R_{G}(x_{1},\Delta,Q)$. 
It can be explained by the fact that  the QCD evolution does not change the difference
$x_{1}-x_{2}$, which is fixed by the kinematics of exclusive reactions.
From Fig.\ \ref{fig:ratnd} one can see that the amount of nuclear  shadowing  for gluons,  given by 
$1-R_{G}(x_{1},\Delta,Q)$, decreases by  approximately a factor of $3.5$   as $Q$ increases from 1.6 to 15 GeV for nuclei with $A=12$, $A=40$, $A=100$, and $A=200$.

Fig.\ \ref{fig:ratx} illustrates increasing 
nuclear shadowing in the limit $x \to 0$. At $Q=5$ GeV, 
when nuclear shadowing is still significant, the ratio $1- R_{G}(x,Q)$ increases by a factor of 10, when $x$ is 
decreased from $x=10^{-2}$ to $x=10^{-4}$ for nuclei with $A=12$, by approximately a factor of 8  for nuclei with  $A=40$, and by approximately a factor of 6 for nuclei with $A=100$ and $A=200$.   

      Figs.\ \ref{fig:ratq}, \ref{fig:ratx}, and  \ref{fig:ratnd}
demonstrate quantitatively that, as $Q$ increases, 
nuclear shadowing decreases, and 
the regime of the generalized color transparency turns into 
the regime of color transparency.

\subsection{The unitarity boundary for the inelastic cross section 
and the phenomenon of perturbative color opacity}  
\label{subsec:Unitarity}

In this review we discuss coherent phenomena with nuclei at small Bjorken $x$. The main tool for the numerical analysis is the DGLAP QCD evolution equation. Then, the question arises: How low in $x$ can one go before the evolution equation becomes unreliable?

One way to approach the answer to this question was presented in Ref.\ \cite{FKS96}. It is based on the observation that the inelastic small-size $q \bar{q}$ dipole-nucleon cross section of Eq.\ (\ref{cs}), being of the leading twist, dominates over the elastic  $q \bar{q}$-nucleon cross section, which is of the next-to-leading twist
\begin{equation}
\sigma^{inel} \gg \sigma^{el} \ .
\label{unit1}  
\end{equation}
When $\sigma^{inel} \approx \sigma^{el}$, the higher twist  expression becomes of the same order as the leading twist term. Hence, the QCD evolution equation becomes inconsistent since it has generated large higher twist terms, which have been neglected in the derivation of the evolution equation. 

This argument demonstrates that $\sigma^{inel}$ given by Eq.\ (\ref{cs})  cannot grow infinitely at small $x$ since that would destroy self-consistency of the DGLAP QCD evolution equation. The upper limit for $\sigma^{inel}$ can be obtained from the constraint imposed by the unitarity of the $S$-matrix for purely QCD interactions and the optical theorem
\cite{FKS96}
\begin{equation}
\sigma^{inel} \leq \frac{\pi r_{N}^2}{(1+\beta^2)}=22 \ {\rm mb} \ ,
\label{unit2} 
\end{equation}
where $r_{N}$=0.8 fm  is the r.m.s. radius of the nucleon,
which is measured in the electroproduction of vector mesons at large $Q^2$
via the $t$-slope of the differential cross section; $\beta \approx 0.4$ is the ratio of the real to imaginary parts of the elastic scattering amplitude.   

In case of inclusive DIS on nuclear targets, the upper limit for the inelastic $q \bar{q}$-nucleus cross section $\sigma^{inel}_{A}$ follows from the unitarity restriction
\begin{equation}
\sigma^{inel}_{A} \leq \frac{1}{2}\sigma^{T}_{A}=\pi R_{A}^2 \ ,
\label{unit3}  
\end{equation}
where $\sigma^{T}_{A}$ is the total $q \bar{q}$-nucleus  cross section. Eq.\ (\ref{unit3}) is applicable for sufficiently heavy nuclei when  $\sigma^{T}_{A}=2\pi R_{A}^2$, where $R_{A}$
is the r.m.s. radius of the nucleus. Using $R_{A}=1.1 A^{1/3}$ fm, one can write Eq.\ (\ref{unit3})  in the form
\begin{equation}
\frac{\sigma^{inel}_{A}}{A} \leq \frac{38 \ {\rm mb}}{A^{1/3}} \ .
\label{unit4}  
\end{equation}

We studied Eqs.\ (\ref{unit2}) and (\ref{unit4}) numerically using the recently developed model for $\sigma^{inel}$ \cite{FGMS99}. The straightforward generalization of Eq.\ (\ref{cs}) for the color triplet ($q \bar{q}$ fluctuation) and color octet ($q \bar{q}g$ fluctuation) dipole scattering cross sections on the nuclear target reads
\begin{eqnarray}
&&\sigma^{q \bar{q}}_{A}(b^2,x^{\prime})
= \frac{\pi^2}{3} b^2 \left[ x^{\prime}
G_{A}(x^{\prime}, \lambda/b^2) \right] \alpha_{s}(\lambda/b^2) \ , \nonumber\\
&&\sigma^{q \bar{q}g}_{A}(b^2,x^{\prime})
= \frac{3\pi^2}{4} b^2 \left[ x^{\prime}
G_{A}(x^{\prime}, \lambda/b^2) \right] \alpha_{s}(\lambda/b^2) \ .
\label{unit5}
\end{eqnarray} 
In our numerical analysis, the nuclear gluon parton density $x^{\prime}G_{A}(x^{\prime}, \lambda/b^2)$ was parameterized as the product of the factor describing shadowing at  $10^{-5} \leq x^{\prime} \leq 0.05$ and antishadowing at $0.05 \div 0.1 \leq x^{\prime} \leq 0.2$ according to Ref.\ \cite{FS99}, and the gluon parton density of the proton. We used the leading order parton parameterization CTEQ4L in  Eqs.\ (\ref{unit2}) and (\ref{unit4}) since the expressions for the  inelastic $q \bar{q}$-nucleon ($q \bar{q}g$-nucleon) and $q \bar{q}$-nucleus ($q \bar{q}g$-nucleus) cross sections have been derived in the leading $\alpha_{s}(Q^2)\,\log(Q^2/\Lambda^2)$ approximation.

The solutions of  Eqs.\ (\ref{unit2}) and (\ref{unit4})
 are presented in Figs.\ \ref{bound1}-\ref{bound4}. For each given $Q_{eff}=\sqrt{\lambda}/b$, the liming value of $x^{\prime}$, which is denoted $x_{lim}$,  at which Eqs.\ (\ref{unit2}) and (\ref{unit4}) are still satisfied, was found. Thus, the kinematics regions allowed by the unitarity restrictions lie to the right of the corresponding curves.

In Fig.\ \ref{bound1}, we plot $x_{lim}$ as a function of $Q^2$ for nuclei with 
$A$=12, 40, 100 and 200, and for the nucleon. One can see that, for nuclei with $A$=200, 100 and 40, the kinematics constraints are more stringent than for the nucleon. Thus, the breakdown of the DGLAP QCD evolution equation 
occurs sooner, i.e. at larger $x$ for a given $Q$, for heavy nuclei than for the proton.

Note that, for heavy enough nuclei, one can select scattering at central impact parameters by taking pairs of nuclei like $^{208}$Pb and $^{40}$Ca.
Since the 
nuclear thickness
at small impact parameters is about a 
factor of 1.5 larger than the average nuclear thickness,
and since the average nuclear thickness is $\propto A^{1/3}$
 the 
plots for $A$=40, 100 and 200
roughly correspond to scattering at the central impact parameters on nuclei with  $A' \approx A/(1.5)^3$.

In Fig.\ \ref{bound2}, we plot $x_{lim}$ as a function of $Q^2$ for nuclei with 
$A$=200. We present three scenarios of gluon shadowing in nuclei. The solid curve corresponds to nuclear shadowing parameterized as in \cite{FS99} (it is the same curve as in Fig.\ \ref{bound1}). The dashed curve corresponds to a smaller amount of shadowing. In this case, gluons are  shadowed as the sea quarks in \cite{FS99}. The dotted curve is for the situation, when nuclear effects are neglected.
In all three cases, at each given $Q$, $x_{lim}$ is larger than the corresponding $x_{lim}$ for the proton target.

Figs.\ \ref{bound3} and \ref{bound4} represent the unitarity boundary for the inelastic $q {\bar q} g$ (color octet) fluctuation scattering cross section on nuclear and nucleon targets, which is given by combining Eqs.\ (\ref{unit4}) and (\ref{unit5}).
 While the trend is quite similar to Figs.\ \ref{bound1} and \ref{bound2}, the unitarity boundary is much more stringent -- $x_{lim}$ is greater by more than the order of magnitude at the corresponding $Q_{eff}$.

The phenomenon of $\sigma^{inel}$ being large at small transverse size $b$ and small $x$ is called color opacity phenomenon. 
The regime of color opacity corresponds to a
small coupling constant $\alpha_{s}(Q^2)$ and large parton densities. Figs.\ \ref{bound1} and \ref{bound2} demonstrate that
it is more advantageous to study 
 the physics of color opacity with heavy nuclei than with the proton. The latter observation can be  explained by the fact that the density of partons per nucleon, even when one takes into account significant nuclear shadowing, is larger for heavy nuclei than for the proton.  

The regime of color opacity is a new field of QCD. 
It
becomes accessible at larger $x$ for heavy nuclei than for the proton target. Hence, it should be feasible to study color opacity at LHC \cite{FKS96,FELIX}, HERA with nuclear beams \cite{FKS96,Krasny}, and eRHIC.

\section{Soft diffractive dissociation on nuclei}
\label{sec:SD}

The present  section is concerned with applications of the principles of color coherence to the domain of soft physics -- diffractive dissociation of hadrons on nuclei and nuclear shadowing in hadron-nuclear collisions. 
Color or cross section fluctuations constitute the basis of a phenomenological description of all the  available data on diffractive dissociation and nuclear shadowing on light and heavy nuclei.

\subsection{Diffractive dissociation as a manifestation of the 
coherence length and color fluctuations}
\label{subsec:DD}

As discussed in 
the Introduction,
one of the principles of color coherence phenomena in QCD is that an energetic hadronic projectile  consists of coherent 
quark-gluon
configurations of very different spatial sizes. These quark-gluon configurations are called color fluctuations. They can  also be called cross section fluctuations since the configurations with different transverse sizes interact with the target with different cross sections. Cross section  fluctuations become significant, when 
the
coherence length $l_{c}$ is large. The importance of the large $l_{c}$  in diffractive dissociation of hadrons was first discovered in \cite{FP}.

It is important to note that one has to distinguish two phenomenological descriptions of strong interactions. In the description, which is more relevant for  hard processes, the color fluctuations are 
quark-gluon
configurations of
small sizes, which  are almost
 fixed. The masses of such fluctuations 
are
undefined. The other description, which is natural for diffractive dissociation, operates with  physical particles of fixed masses. Then, it is 
more appropriate to consider the distribution over transverse sizes.
Using the principles of color coherence, one can naturally relate the two descriptions of strong interactions.

Let us consider  diffractive dissociation of a hadron with   mass $m$ and  large momentum $p_{lab}$ on   a hadronic target $T$
in the laboratory reference frame. The 
incoming hadron can fluctuate into a hadronic state with mass $M^{\ast}$
due to the uncertainty principle. 
 The coherence length $l_{c}$ 
corresponds to
the time, during which the incoming hadron stays in the state with mass $M^{\ast}$,  
\begin{equation}
l_{c} = \frac{1}{\Delta E}=
\left(\sqrt{M^{\ast\,2}+p_{lab}^2}-\sqrt{m^2+p_{lab}^2}\right)^{-1} 
\simeq \frac{2p_{lab}}{M^{\ast 2}-m^2} \ .
\label{cls}
\end{equation}
From Eq.\ (\ref{cls}), one can immediately see that the coherence length grows with the energy of the incoming hadron (which is proportional to $p_{lab}$).

If the coherence length is greater than the typical size of the target $2R_{T}$, 
\begin{equation} 
l_{c} \geq 2R_{T} \ ,
\label{cond2} 
\end{equation}
then all the fluctuations, whose masses satisfy Eqs.\ (\ref{cls}) and (\ref{cond2}), interact coherently with the entire target. 
This fact 
has three consequences. Firstly, as energy grows, more  diffractive states of ever-increasing mass can be produced. Secondly, the large $l_{c}$ corresponds to small momentum 
transfers
to the target. It means that the target (nucleon or nucleus) has a large probability to remain in its ground state while the interaction takes place. Moreover, if the target is a nucleus, the suppression of the scattering amplitude due to the nuclear form factor is small. The first and second consequences  manifest themselves in a large probability of coherent diffractive processes on nuclei.
Thirdly, 
the coherence of the fluctuations means that  they can be considered as frozen during the interaction with the target. This enables one to introduce 
a function $P(\sigma)$, which gives 
the probability to find a hadronic (quark-gluon) configuration, which interacts with the target with the cross section $\sigma$, in the energetic projectile.  $P(\sigma)$
can be most naturally introduced  within the framework of  the basis of states, which consists of configurations of definite scattering cross sections  with the target \cite{GW}.

The vector of state of an energetic incident hadron $|\Psi \rangle$ can be represented as a coherent superposition of eigenstates $|\Psi_{k} \rangle$  of the scattering matrix   
\begin{equation}
|\Psi \rangle=\sum_{k}c_{k}|\Psi_{k} \rangle \ ,
\label{gw1}
\end{equation}
where
\begin{eqnarray}
Im T|\Psi_{k} \rangle&=&t_{k}  |\Psi_{k} \rangle \ , \nonumber\\
\sum_{k}|c_{k}|^2&=&1 \ .
\label{gw2}
\end{eqnarray}

Here, $T$ is the scattering operator, and $t_{k}$ is the imaginary part of the  eigenvalue corresponding to the eigenstate $|\Psi_{k} \rangle$.

Since various states $|\Psi_{k} \rangle$ interact with the target with different cross sections $\sigma_{k}$, which, by the optical theorem, are related to the imaginary part of the scattering amplitude $t_{k}$,
\begin{equation}
\sigma_{k}=t_{k} \ ,
\label{gw3}
\end{equation}
the coherent superposition of the eigenstates, which forms  the final state 
emerging after the scattering, could be 
different from the initial state. Thus, the formalism
of eigenstates of the scattering matrix is natural for
 describing diffractive dissociation of hadrons.

Using Eqs.\ (\ref{gw1})-(\ref{gw3}), diffractive dissociation can be presented as follows. 
Diffractive scattering occurs when the final state carries the same quantum numbers as the initial state, i.e. whenever the initial state overlaps any $|\Psi_{k} \rangle$. Then,
the total diffractive differential cross section at $t=0$ can be presented  as 
\begin{equation}
\Big(\frac{d \sigma}{dt} \Big)^{diff}_{t=0}=\frac{1}{16\pi}\sum_{k}|\langle\Psi_{k}| Im T |\Psi \rangle|^2=\frac{1}{16\pi}\sum_{k}|c_{k}|^2 t_{k}^2 \equiv \frac{1}{16\pi}\langle \sigma^2 \rangle \ .
\label{o1}
\end{equation}

In Eq.\ (\ref{o1}),  we have used the completeness of the set of states $|\Psi_{k}\rangle$ and the optical theorem\ (\ref{gw3}). We have also introduced  the second moment $\langle \sigma^2 \rangle$ of the distribution over cross sections $P(\sigma)$.
Similarly, the elastic diffractive differential cross section at $t=0$ reads
\begin{equation}
\Big(\frac{d \sigma}{dt} \Big)^{el}_{t=0}=\frac{1}{16\pi}|\langle\Psi| Im T |\Psi \rangle|^2=\frac{1}{16\pi}\Big(\sum_{k}|c_{k}|^2 t_{k}\Big)^2 \equiv \frac{1}{16\pi}\langle \sigma \rangle  ^2 \ ,
\label{o2}
\end{equation}
where $\langle \sigma \rangle$ is the first moment of $P(\sigma)$. 

Subtracting the elastic cross section 
from the total diffractive cross section, one obtains 
the diffractive  dissociation cross section  (inelastic diffractive cross section)
\begin{equation}
\Big(\frac{d \sigma}{dt} \Big)^{inel}_{t=0}=\Big(\frac{d \sigma}{dt} \Big)^{diff}_{t=0}-\Big(\frac{d \sigma}{dt} \Big)^{el}_{t=0}=\frac{1}{16\pi}\Big(\langle \sigma^2 \rangle-\langle \sigma \rangle ^2\Big) \ .
\label{ineldiff}
\end{equation}

Eq.\ (\ref{ineldiff}) was first derived in \cite{MP78} in order to describe diffractive dissociation within the framework of cross section fluctuations. Eq.\ (\ref{ineldiff})
explicitly demonstrates
 that diffractive dissociation occurs only if different components $|\Psi_{k} \rangle$ of the incident hadron interact with the target with different strengths $t_{k}$, i.e. when cross section fluctuations take place  in the wave function of the incoming hadron.

\subsection{Properties of $P(\sigma)$}
\label{subsec:PP}

 The formalism of scattering matrix eigenstates provides an economic way 
to take into account the composite structure of energetic  projectiles.  
Due to the frozenness condition,  each state $|\Psi_{k}\rangle$
 scatters independently on the target and then contributes coherently to the final  state with the probability $|c_{k}|^2$, see also Eq.\ (\ref{gw1}). Then, one can
introduce the distribution over cross sections $P(\sigma)$ \cite{rev1}
\begin{equation}
P(\sigma)=\sum_{k}|c_{k}|^2 \delta(\sigma-\sigma_{k}) \ .
\label{ll}
\end{equation} 
One can see from Eq.\ (\ref{ll}) that  
it is the composite nature of hadrons that gives rise  to the term 
cross section fluctuations. 

The cross section distribution function  $P(\sigma)$ comprises perturbative 
and non-perturbative information on the wave function of the hadron. The modelling of $P(\sigma)$  involves  pQCD calculations, valid at   small $\sigma$,    and constraints on first moments of $\sigma$,  extracted from experiments. In addition, it is required that
$P(\sigma) \to 0$ at $\sigma \to \infty$.

The quark counting rule fixes the low-$\sigma$ behaviour of $P_{h}(\sigma)$ for hadrons \cite{BBFHS}
\begin{equation}
P_{h}(\sigma) \propto \sigma^{n_{q}-2} \ ,
\label{o3}
\end{equation}
where $n_{q}$ is the number of valence quarks in the hadron. Thus, for protons, when $\sigma$ is much smaller than the average value of $\sigma$, i.e. when $\sigma \ll \langle \sigma \rangle$, one obtains that 
\begin{equation}
P_{p}(\sigma) \sim \sigma  \ .
\label{pproton}
\end{equation}

For pions, the quark counting rule predicts at $\sigma \ll \langle \sigma \rangle$ \cite{BBFHS}
\begin{equation}
P_{\pi}(\sigma) \sim {\rm const} \ .
\label{ppion11}
\end{equation}

However,  one can go further than the simple counting rule in case of pions. In order to find  $P_{\pi}(\sigma)$ at $\sigma \ll \langle \sigma \rangle$,  one can apply the QCD factorization theorem \cite{rev1,BBFHS} and find the constant entering Eq.\ (\ref{ppion11}):
\begin{equation}
P_{\pi}(\sigma)=\frac{6f^2_{\pi}}{5 \alpha_s(4k^{2}_{t})\bar x
  G_{N}(\bar x, 4k^{2}_{t})} \ .
\label{ppion}
\end{equation}
In this equation, $\alpha_{s}(4k^{2}_{t})$ is the QCD running coupling constant;
$G_N(\bar x,4k^{2}_{t})$ is the gluon distribution in the nucleon;
 $\bar x= 4k^2_t/s_{\pi N}$ , where $s_{\pi N}$ is the center of mass energy squared; $k^2_t \propto 1/b^2$, 
where $b$ is the transverse size of the $q \bar{q}$ pair of the wave function of the pion; $f_{\pi}$ is the constant for the $\pi \to \mu \nu$ decay. 

Note that the calculations of the cross section  of exclusive electroproduction of vector mesons \cite{Brod94} and that of  $P_{\pi}(\sigma)$ \cite{rev1,BBFHS} use a similar technique. Therefore, the confirmation of the results of \cite{Brod94} by the ZEUS data at HERA \cite{HERA} provides an indirect confirmation of the calculation of $P_{\pi}(\sigma \ll \langle \sigma \rangle)$.

Using the  
QCD factorization theorem, one can compute $P_{\gamma}(\sigma)$ for photons at $\sigma \ll \langle \sigma \rangle$. It was found that the behaviour of $P_{\gamma}(\sigma)$ at small $\sigma$ is divergent \cite{FRadS98}
\begin{equation}
P_{\gamma}(\sigma) \propto \frac{1}{\sigma} \ .
\end{equation}
Such a  behaviour is consistent with the quark counting rule of Eq.\ (\ref{o1}). Since the quark and antiquark in the photon wave function are created at the same point, they have a large probability to stay close to each other.

 Suggested parameterizations of  the proton and pion distribution 
functions $P(\sigma)$  describe correctly the above discussed  behaviour at small and large $\sigma$ \cite{BBFHS}
\begin{eqnarray}
P_{p}(\sigma)&=&N(a,n)\frac{\sigma / \sigma_{0}}{\sigma / 
\sigma_{0}+a}e^{-(\sigma-\sigma_{0})^{n}/(\Omega \sigma_{0})^{n}} \ , \nonumber\\
P_{\pi}(\sigma)&=&N(a,n)e^{-(\sigma-\sigma_{0})^{n}/
(\Omega \sigma_{0})^{n}} \ . 
\label{param}
\end{eqnarray} 
$P_{p}(\sigma)$ and $P_{\pi}(\sigma)$ as functions of $\sigma$ are presented in Fig.\ \ref{fig4} \cite{BBFHS}.

The free parameters of the parameterizations\ (\ref{param}) are chosen 
in order to satisfy the following constraints on 
the first few  moments of the distribution $P(\sigma)$:
\begin{eqnarray}
&\int& P(\sigma) d\sigma = 1 \ , \nonumber\\
&\int& P(\sigma) \sigma  d\sigma = \langle \sigma \rangle \ , \nonumber\\
&\int& P(\sigma) \sigma^2 d\sigma =\langle \sigma^2 \rangle=(1+\omega_{\sigma}) 
\langle \sigma \rangle ^2 \ , \nonumber\\
&\int& P(\sigma) (\sigma - \langle \sigma \rangle)^3 d \sigma \approx 0 \ .
\label{moments}
\end{eqnarray}
The first equation  is the normalization of $P(\sigma)$. The first
moment $\langle \sigma \rangle$  is the total hadron-nucleon cross 
section. 

The third of Eqs.\ (\ref{moments}) is 
a relation between the  second moment $\langle \sigma^2 \rangle$
and the quantity $\omega_{\sigma}$. Using Eq.\ (\ref{ineldiff}), one can write 
that 
\begin{equation}
\omega_{\sigma}=\frac{\Big(\frac{d \sigma}{dt} \Big)^{inel}_{t=0}}{\Big(\frac{d \sigma}{dt} \Big)^{el}_{t=0}}=\frac{\langle \sigma^2 \rangle - \langle \sigma \rangle ^2}{\langle \sigma \rangle ^2} \ ,
\label{omega}
\end{equation} 
which demonstrates that $\omega_{\sigma}$ is proportional to  the amount of diffractive dissociation.
$\omega_{\sigma}$ can be extracted from the experiment.
From the  data on the total cross section 
of neutron-deuteron scattering \cite{nD} and from the theoretical analysis of
\cite{pD}, which used the  
proton-deuteron data \cite{piD},
$\omega_{\sigma}$=0.25 for $p_{lab} \approx$ 200 GeV/c. Another source of information on $\omega_{\sigma}$ is the data on diffractive dissociation on hydrogen. The data  leads to a similar value of $\omega_{\sigma}$ \cite{BBFHS}. 

The last equation in Eq.\ (\ref{moments}) is based on  
the analysis of \cite{BBFHS} of 
the data on coherent 
diffractive dissociation of protons on deuteron \cite{pDdiff}. It 
indicates  that $P_{p}(\sigma)$ is symmetrical around its average value. 
In order to impose further restriction of higher moments of $P(\sigma)$, one has to consider diffractive dissociation on heavier nuclei, for example, on $^4$He.

\subsection{Total and diffractive dissociation cross sections of hadron-deuteron scattering}
\label{subsec:deuteron}

In the previous subsection we demonstrated that
in order to study the second moment of  $P(\sigma)$, $\langle \sigma^2 \rangle$, one can study either diffractive dissociation on hydrogen (see Eq.\ (\ref{ineldiff})) or the total hadron-deuteron cross section. In this subsection, the latter approach is considered in detail.
 
As discussed in Subsec.\ \ref{subsec:shadowing}, there is a deep connection between high energy diffraction of hadrons on the proton and nuclear shadowing in the total hadron-nucleus cross section.
Using  the space-time picture of high energy scattering, 
which follows from the quantum field theory,
 V. Gribov has derived  the total cross section of hadron-deuteron
scattering $\sigma_{tot}^{hD}$ \cite{Gribov4} (see also Eq.\ (\ref{shdeu}))
\begin{equation}
\sigma_{tot}^{hD}=\sigma_{tot}^{hp}+\sigma_{tot}^{hn}-\frac{1}{4\pi}\int
dt
S(4t) \Big(\int dM^2 \frac{d^2 \sigma^{h + p \to X + p}}{dM^2 dt}
  +
\frac{d \sigma^{h + p \to h + p}}{ dt} \Big) \ ,
\label{deu1}
\end{equation}
where $\sigma_{tot}^{hp}$ and $\sigma_{tot}^{hn}$ are the total hadron-proton and hadron-neutron cross sections;
$d^2 \sigma / dM^2 dt$ is the cross section for producing a diffractive  state
$X$ with mass $M$ in the reaction $h + N \to X + N$; $t$ is the squared four-momentum transfer; 
$S(t)$ is the
electro-magnetic form factor of the 
deuteron.  For simplicity, spin effects in the wave function
of the deuteron and the real part of the diffractive scattering amplitude have been neglected.

In spite of different space-time 
evolution pictures of scattering processes in quantum field theory and
 in  non-relativistic quantum mechanics, Eq.\ (\ref{deu1}) generalizes the earlier result of  \cite{Glauber}, when only the second term 
in the bracket
of Eq.\ (\ref{deu1}) (elastic shadowing) was taken into account.

Eq.\ (\ref{deu1}) relates the amount of nuclear shadowing in the total cross section of hadron-deuteron scattering to the cross section of  diffractive dissociation of the hadron on the proton.
  Nuclear shadowing due to  diffractive dissociation (inelastic diffraction) is called inelastic shadowing because it is
caused by the excitation and 
propagation of inelastic intermediate states. Fig.\ \ref{fig:str1} depicts that the inelastic shadowing term appears due to the interference between the diffractive scattering of the incident hadron off the proton and the neutron of the deuterium target. 
Although inelastic shadowing corrections are rather small for hadron-nucleus collisions, inelastic shadowing  is the only mechanism, which relates nuclear shadowing in inclusive  DIS on nuclei  to diffraction in DIS on proton.

Within the  formalism of cross section fluctuations,  Eq.\ (\ref{deu1})
was rederived  in \cite{KL},  
\begin{equation}
\sigma_{tot}^{hD}=2 \langle \sigma \rangle - \frac{\langle \sigma^2 
\rangle}{8 \pi ^2} \int e^{-q^2 \beta}S(q) d^2 q  \ .
\label{deu2}
\end{equation}
Here, it is assumed that $\sigma_{tot}^{p}=\sigma_{tot}^{n}=\langle \sigma \rangle$ and that the spin effects in the deuteron wave function and 
the real part of the diffractive scattering amplitude are negligible. Eq.\ (\ref{deu2})  explicitly demonstrates how $\langle \sigma^2 \rangle$ and $\omega_{\sigma}$  can be extracted from the shadowing term in the total hadron-deuteron cross section. Using the data on elastic and inelastic 
shadowing in $\sigma_{tot}^{nD}$ \cite{nD} and  $\sigma_{tot}^{pD}$ \cite {piD} (see also \cite{pD}), one finds that $\omega_{\sigma}(p,n)=0.25$ for the average values of $p_{lab}$, $p_{lab}$=200 GeV/c.

For a  pion projectile, the value of  
$\omega_{\sigma}(\pi)$=0.4 at $p_{lab}$=200 GeV/c was extracted in \cite{BBFS2} from 
the data on $\pi D$ scattering of \cite{piD}.
This value of $\omega_{\sigma}$  is in accord with the theoretical expectations, ensuing  
from 
the quark counting rule and the triple Pomeron exchange model. 
The larger value of $\omega_{\sigma}(\pi)$ for pions than for protons 
means that the distribution over cross
sections $P(\sigma)$ is broader for pions than that for
protons. This fact is  reflected in the parameterizations \ (\ref{param}).

A direct comparison of Eqs.\ (\ref{omega}), (\ref{deu1}) and (\ref{deu2}) enables one to identify $\omega_{\sigma}$ as the ratio of the diffractive dissociation to elastic cross sections
\begin{equation}
\omega_{\sigma}=\frac{\langle \sigma^2 \rangle - \langle \sigma \rangle ^2}{\langle \sigma \rangle ^2}=\Big(\int \int S(4t) \frac{d^2 \sigma^{h + p \to X + p}}{dM^2 dt}dM^2dt \Big)\Big/ \Big(\int S(4t)
\frac{d \sigma^{h + p \to h + p}}{ dt}dt \Big) \ .
\label{deu4}
\end{equation}
Eq.\ (\ref{deu4})  explicitly demonstrates the connection
between diffractive dissociation, inelastic nuclear shadowing, and  cross section fluctuations.

The third moment $\langle \sigma^3 \rangle$ of $P_{h}(\sigma)$
can be found by considering diffractive dissociation of the hadron $h$ on deuteron. The reaction $p + D \to X + D$ was analyzed in \cite{BBFHS}, where the ratio of the proton coherent diffractive dissociation cross sections on deuterons 
and hydrogen was presented as
\begin{equation}
\Big(\frac{d\sigma_{diff}}{dt}\Big)^{pD}_{t=0} \Big /
\Big(\frac{d\sigma_{diff}}{dt}\Big)^{pp}_{t=0}=4-\frac{1}{2\pi^2}\frac{\langle
\sigma^3 \rangle -\langle \sigma \rangle \langle \sigma^2
\rangle}{\langle \sigma^2 \rangle - 
\langle \sigma \rangle^2}\int d^2 p_{\perp}S(p_{\perp}) \ .
\label{deu5}
\end{equation}
Using Eq.\ (\ref{deu5}) and  the data on coherent diffraction of protons on deuterium \cite{pDdiff}, one arrives at  
 $\int P_{p}(\sigma)(\sigma - \langle \sigma \rangle)^3 d \sigma \approx 0$ 
(see \cite{BBFHS} for details).

\subsection{Diffractive dissociation on He$^4$}
\label{DHe}

In order to study further the importance of color fluctuations in energetic hadrons and  higher moments of the distribution $P(\sigma)$, one
should consider heavier nuclei. The only existing experimental data on
inclusive coherent diffraction on light nuclei is the data on proton
diffractive 
dissociation on $^4$He in the reaction $p + ^{4}{\rm He} \to X + ^{4}{\rm He}$ 
\cite{He-4}.  

Using the cross section (color) fluctuation formalism along with the Gribov-Glauber multiple scattering formalism \cite{Gribov4,Glauber}, the ratio of diffractive
differential cross section on $^4$He and on the proton at $t=0$ can be presented as a series over moments of $P_{p}(\sigma)$ \cite{Guzey}
\begin{eqnarray}
\Bigg(\frac{d\sigma_{diff}}{dt}\Bigg)^{pHe}_{t=0} \Bigg/
 \Bigg(\frac{d\sigma_{diff}}{dt}\Bigg)^{pp}_{t=0}&=
&16-\frac{6\gamma}{\omega_{\sigma}}\Bigg(\frac{\langle\sigma^3\rangle}
{\langle\sigma\rangle^3}-\frac{\langle\sigma^2\rangle}{\langle\sigma
\rangle^2}\Bigg)+\frac{59\gamma^2}{48\omega_{\sigma}}\Bigg(\frac{\langle
\sigma^4\rangle}{\langle\sigma\rangle^4}-\frac{27}{59}\frac{\langle
\sigma^2\rangle^2}{\langle\sigma\rangle^4}-\frac{32}{59}\frac{\langle
\sigma^3\rangle}{\langle\sigma\rangle^3}\Bigg) \nonumber\\     
  &-&\frac{5\gamma^3}{32\omega_{\sigma}}\Bigg(\frac{\langle
\sigma^5\rangle}{\langle\sigma\rangle^5}-\frac{4}{5}\frac{\langle\sigma^3
\rangle\langle\sigma^2\rangle}{\langle\sigma\rangle^5}-\frac{1}{5}
\frac{\langle\sigma^4\rangle}{\langle\sigma\rangle^4}\Bigg) \ .
\label{hel1}
\end{eqnarray}

Here, $\langle \sigma \rangle$=41 mb is the total hadron-nucleon cross
section at $p_{lab}=200$ GeV/c. The parameter $\gamma$, defined as $\gamma=\langle \sigma \rangle / \pi(\alpha+\beta)$, emerges from the integration over transverse momenta of the proton-nucleon
 scattering amplitudes with the $^4$He ground state wave function; $\alpha$=23 (GeV/c)$^{-2}$ is the slope of the ground state wave function of $^4$He, which is parameterized in the exponential form \cite{LS}; $\beta$=10.5 (GeV/c)$^{-2}$ is the slope of hadron-nucleon cross section, which is taken as the average of the slopes of the  elastic and diffractive cross sections \cite{Guzey}. 
This yields
$\gamma$=1.01.

Then, using the parameterization \ (\ref{param}) in order to compute the higher moments of $P_{p}(\sigma)$, Eq.\ (\ref{hel1}) leads to \cite{Guzey}
\begin{equation}
r=\Big( \frac{d\sigma_{diff}}{dt} \Big)^{pHe}_{t=0}/ \Big( \frac{d\sigma_{diff}}{dt} \Big)^{pp}_{t=0}=6.8 \div 7.6 \ . 
\label{nnnew1}
\end{equation}
The lower value of $r$ corresponds to $n=2$ in Eq.\ (\ref{param}), and the upper value of $r$ corresponds to $n=6$ and $n=10$ in Eq.\ (\ref{param}).

Eq.\ (\ref{nnnew1}) is  in an excellent agreement with the experimental data on proton coherent diffractive dissociation on helium \cite{He-4}. 
Integrating the data over the available range of diffractive masses $2.5 \leq M^2 \leq 8$ GeV$^2$, it was found in \cite{Guzey} that
\begin{equation}
r^{exp}=7.1 \pm 0.7  \ .
\end{equation}
Here, the main error originates from the extrapolation of the data to $t=0$.
It is important to note that, although the comparison to the theoretical value of $r$
implies the integration over all diffractive masses $M^2$, the
correction, related 
to the inclusion of non-measured
small  diffractive masses, is small \cite{Guzey}.

The remarkable agreement between the theory and experiment demonstrates that, indeed, significant color fluctuations (cross section fluctuations) are present in energetic protons. The phenomenological description of those fluctuations  by Eq.\ (\ref{param}), which is based on the notion of color fluctuations, successfully reproduces the existing experimental data on diffraction on helium and heavier nuclei (see Subsect.\ {\ref{subsec:DHN}) with an a
accuracy of 10\% without introducing any new free parameters. The latter fact is a sign of universality  of the description that uses the distribution over cross sections $P(\sigma)$.

One can also compare  the theoretical prediction with the data on the total diffractive cross section of proton-$^4$He scattering. Combining the Gribov-Glauber multiple scattering formalism,  cross section fluctuations and the parameterizations\ (\ref{param}), the total coherent diffractive cross section of proton-$^4$He scattering $\sigma^{diff}_{He-4}$  was obtained in \cite{Guzey2} 
\begin{eqnarray}
&&\sigma^{diff}_{He-4}=3.5 \ {\rm mb}, \  n=2 \ , \nonumber\\
&&\sigma^{diff}_{He-4}=3.7 \ {\rm mb}, \  n=6, 10 \ .
\label{th1}
\end{eqnarray}

The result of Eq.\ (\ref{th1}) is in a good agreement with the experimental data of \cite{He-4}, integrated over $0.04 
\mbox{ GeV}^2
\leq |t| \leq 0.4$ GeV$^2$ and over diffractive masses  $0 
\mbox{ GeV}^2
\leq M^{2} \leq 10$ GeV$^2$:
\begin{equation}
\sigma^{diff}_{He-4}(exp)=2.6 \pm 0.9 \ {\rm mb} \ .
\label{th2}
\end{equation}
In Eq.\ (\ref{th2}), the error is dominated by the uncertainty of the extrapolation of the experimental data into the unmeasured regions of small $t \leq 0.04$ GeV$^2$ and small diffractive masses $M^2 \leq 1.2$ GeV$^2$.

Fig.\ \ref{fig5} presents a comparison of the theoretical predictions for the total cross sections of coherent diffraction of protons and neutrons on nuclei to the available  experimental data. The value of $\sigma^{diff}_{He-4}(exp)$, given by Eq.\ (\ref{th2}), is presented as a full circle. The theoretical prediction for $\sigma^{diff}_{He-4}$, given by Eq.\ (\ref{th1}), is presented by two dashed curves that correspond to $n=2$ (lower curve) and $n=6$ and $n=10$ (upper curve). 
 
\subsection{Diffractive dissociation on heavy nuclei}
\label{subsec:DHN}

Cross section (color) fluctuation formalism can also be successfully applied to  diffractive dissociation on heavy nuclei. 
For heavy nuclei, one can neglect the slope of the elementary hadron-nucleon elastic scattering amplitude as compared to the slope of the nuclear wave function as well as correlations in the nuclear wave function. Then, in this approximation, the coherent  diffractive dissociation cross section $\sigma_{diff}^{hA}$ of hadrons $|h \rangle$ on the nucleus with the atomic number $A$  can be presented as \cite{FMS}
\begin{equation}
\sigma_{diff}^{hA}=\int d^2 b \Bigg(\int d \sigma P_{h}(\sigma)\sum_{n} |\langle h| F(\sigma,b)|n \rangle|^2-\Big(\int d \sigma P_{h}(\sigma) |\langle h| F(\sigma,b)|h \rangle| \Big)^2 \Bigg) \ ,
\label{Adiff}
\end{equation} 
where $b$ is the impact parameter; $|n \rangle$ is the diffractively produced final state; $F(\sigma,b)$ is the Glauber profile function, defined 
as
$F(\sigma,b)=1-e^{-\sigma T(b) /2}$, where $T(b)=\int^{\infty}_{-\infty} \rho_{A}(b,z) dz$, $\rho_{A}(b,z)$ is the nuclear density.
Using completeness of the states $|n \rangle$, $\sigma_{diff}^{hA}$ can be presented as
\begin{equation}
\sigma_{diff}^{hA}=\int d^2 b \Bigg(\int d \sigma P_{h}(\sigma) |\langle h| F^{2}(\sigma,b)|h \rangle|-\Big(\int d \sigma P(\sigma) |\langle h| F(\sigma,b)|h \rangle| \Big)^2 \Bigg) \ .
\label{sdiff}
\end{equation}

The theoretical prediction of Eq.\ (\ref{sdiff}) for protons (neutrons) is compared to the experimental data in Fig.\ \ref{fig5}, where $\sigma_{diff}^{pA}$ is presented as a function of $A$. The three solid curves correspond to the calculation with $n$=2, 6
and 10 of the parameterization\ (\ref{param}).
The $A$ dependence of $\sigma_{diff}^{pA}$
is of the approximate form $A^{0.8}$ for $A \approx 16$ and $A^{0.4}$ for $A \approx 200$.  
There are only two sets of experimental data on coherent nuclear diffraction of nucleons on heavy nuclei -- the data  on $n + A \to p \pi^{-} + A$ for the mass interval $1.35 \leq M \leq 1.45$ GeV \cite{Zielinski} and the data for emulsion targets  \cite{Emulsions}. In Fig.\ \ref{fig5}, the data of \cite{Zielinski}  is presented as stars and the data of \cite{Emulsions} is presented as triangles.

As one can see from Fig.\ \ref{fig5},  the theoretical prediction of the $A$ dependence of coherent diffraction of protons and neutrons  on nuclei, based on color fluctuations accumulated in $P_{p}(\sigma)$, reproduces the experimental data well. 
Note that the data points for the reaction $n + A \to p \pi^{-} + A$ lie systematically below the theoretical (solid) curves because the data points represent  only one channel (a particular final state) of the total diffractive cross section.

For pion projectiles, the approach, which uses $P_{\pi}(\sigma)$, also describes the available experimental  data well \cite{FMS}.

Inelastic diffraction (diffractive dissociation) on nuclei arises due to scattering at impact parameters that are rather close to the nuclear surface, where the nucleus is not so opaque and the nuclear density is small. The dominant contribution to inelastic diffraction comes from the fluctuations of the projectile, whose size is close to the average size of the projectile. This fact can be seen from the approximate diffractive cross section $\sigma^{appr\,hA}_{diff}$, which was computed  by expanding $F(\sigma,b)$ about $\sigma=\langle \sigma \rangle$ in Eq. (\ref{Adiff}) \cite{FMS}
\begin{equation}
\sigma^{appr\,hA}_{diff}=\frac{\omega_{\sigma}\langle \sigma \rangle^2}{4} \int d^2 b \, T(b)^2 \, e^{-\langle \sigma \rangle T(b)} \ .
\label{appr1}
\end{equation}
Eq.\ (\ref{appr1}) is quantitatively accurate for $A \leq 50$ and qualitatively good for all $A$. It demonstrates that the $A$ dependence is mainly determined by the value of $\langle \sigma \rangle$. For  $A \geq 50$, Eq.\ (\ref{appr1})  gives a weaker $A$ dependence than the  exact formula (\ref{Adiff}). 
The deviation at large $A$ should come from configurations 
with
relatively small $\sigma$.

It is important to note that 
the numerical analysis of \cite{FMS} shows that it is impossible to study the dominance of small size contributions in reactions of coherent nuclear diffraction simply by increasing $A$, because it requires unreasonably heavy nuclei. Consequently, one should explore the domain of hard diffraction in the search for the dominance of small-size configurations.   

\subsection{Cross section fluctuations in heavy-ion collisions}
\label{sec:Heavy ion}

Color fluctuations also play  an important role in inelastic reactions with nuclei. The effect of color fluctuations in the wave function of hadrons on multiplicity of binary nucleon-nucleon collisions and the transverse energy production  in central high-energy nuclear collisions was calculated in \cite{HBBFS}. It was  found that the color fluctuations account for the large  multiplicity of binary nucleon-nucleon collisions and the transverse energy fluctuations found experimentally.
The fluctuation
parameter $\omega_{\sigma}$ (see Eq.\ (\ref{omega})) increases at $\sqrt{s} \leq 60$ GeV 
and decreases for
 $\sqrt{s}\geq 400$ GeV. Thus, $\omega_{\sigma}$ is likely to be close to its maximum, 
$\omega_{\sigma} \sim 0.3 \div 0.35$,
for the energies to 
be studied in $AA$ collisions at RHIC.
 Therefore, the effects of color fluctuations in the hadron wave function are relevant for the RHIC program and
should be included in the event generators of central 
nucleus-nucleus
collisions.

\section{Concluding remarks}
\label{sec:Conclusions}

We find that QCD unambiguously predicts the onset of color transparency  in 
hard coherent meson electroproduction on nuclei at small Bjorken $x$ and $Q \to \infty$. Thus, coherent hard phenomena are now becoming  an
effective instrument in the investigation of the
 quark-gluon structure of hadrons and nuclei.

At the moment, there are many uncertainties in the subject  of extremely small Bjorken $x$ phenomena, where the QCD evolution 
equation becomes inapplicable.
 However,
the conclusion that the onset of the new QCD regime of 
large parton densities can be achieved in $e\,A$ collisions at HERA and 
$p\,p$ collisions at LHC, seems to be model-independent.

\section{Acknowledgements}

We thank our collaborators
S.~Brodsky, J.~Collins,  A.~Freund, W.~Koepf,
G.A.~Miller, A.H.~Mueller, M.~McDermott, and 
A.~Radyushkin
who have contributed to many of the analyses discussed in the review.
We are indebted to N.~Auerbach for fruitful discussions of 
isobaric excitations in electroweak transitions,
and to  L.~Gerland for valuable comments.
V.G. would like to thank J.~Collins for providing the CTEQ 
QCD evolution package. 

This work was supported by the U.S. Department of Energy grant number 
DE-FG02-93ER-40771,
the Israeli Academy of Science, and the Australian Research Council. 




\begin{figure}
\epsfig{file=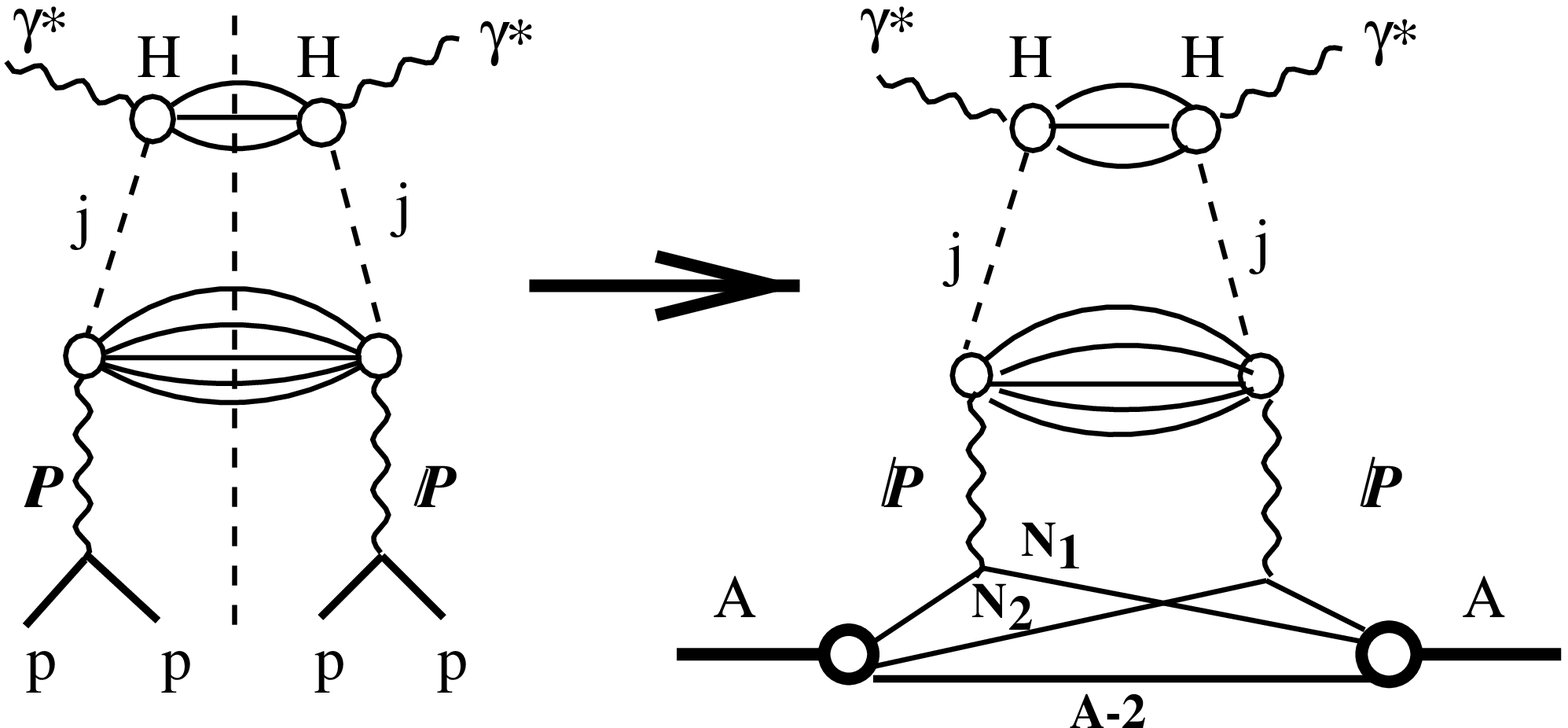,height=8cm,width=14cm}

\vspace{1cm}
\caption{Diagrams demonstrating the relationship between DIS diffraction on protons in the reaction $\gamma^{\ast}+p \to X+p^{\prime}$ and the leading contribution to nuclear shadowing in inclusive DIS on nuclei.}
\label{fig:str1}
\end{figure}



\begin{figure}
\epsfig{file=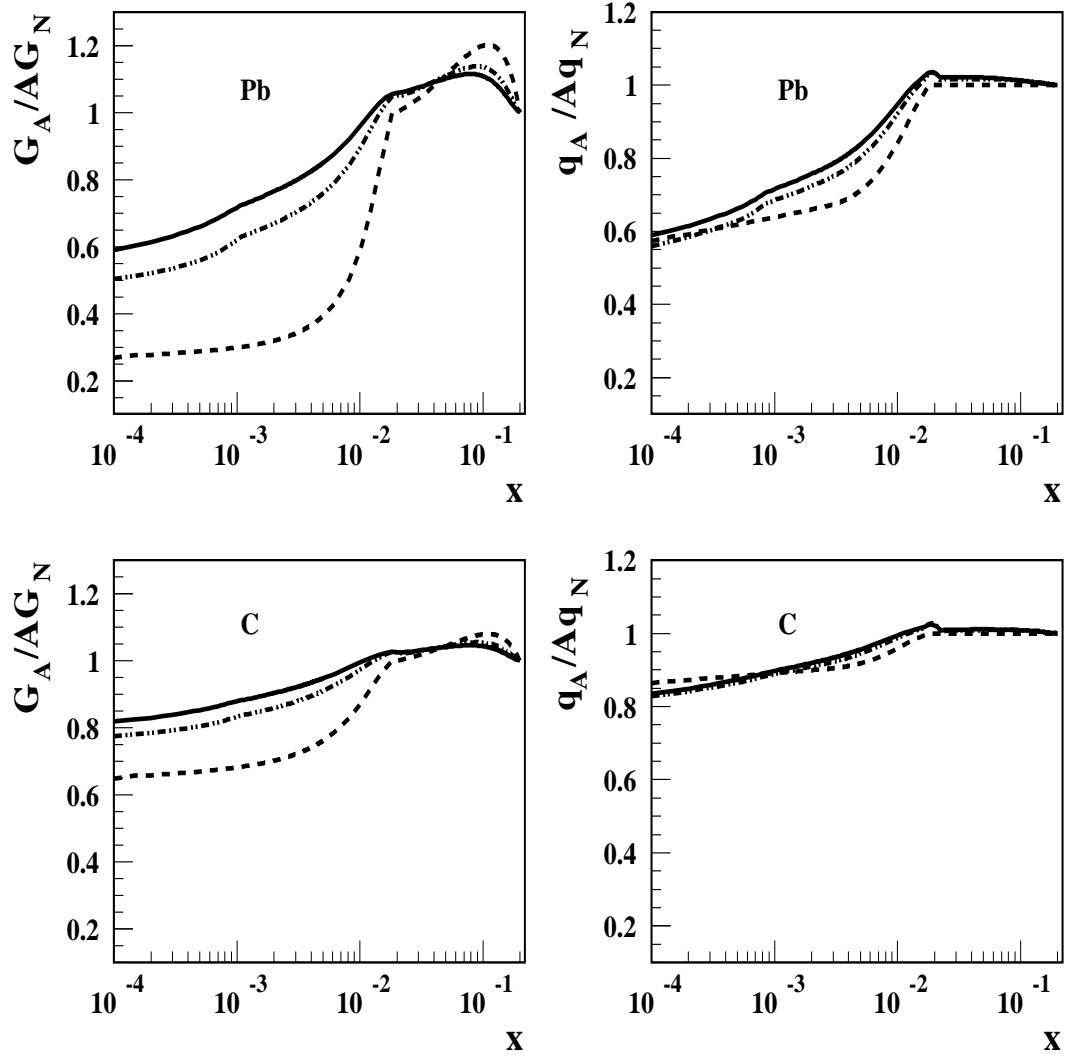,height=14cm,width=14cm}
\vspace{1cm}

\caption{The ratios of the nuclear and nucleon gluon parton densities $G_{A}/AG_{N}$ and the quark parton densities $q_{A}/Aq_{N}$ for lead (Pb) and carbon (C) as a function of $x$. The dotted, dashed and solid curves correspond to $Q$=2, 5 and 10 GeV, respectively. 
}
\label{fig:shh}
\end{figure}


\begin{figure}
\epsfig{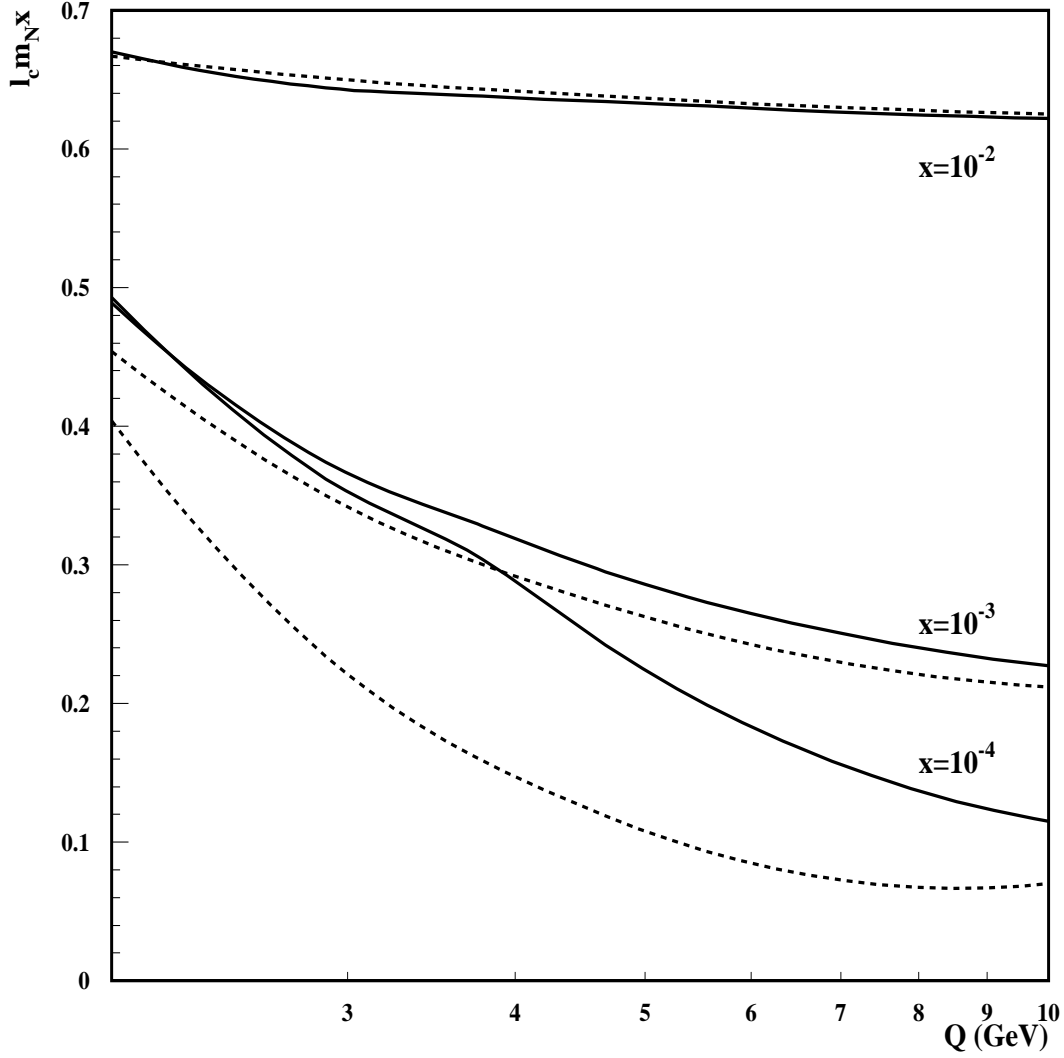}
\vspace{1cm}

\caption{$l_{c}m_{N}x$  as a function of $Q$ at $x$=10$^{-4}$, 10$^{-3}$, and 10$^{-2}$ as given by Eqs.\ (\ref{cl3}) and (\ref{beta2}). The solid curves correspond to the parameterization of the diffractive parton densities of\ [31]. 
The results, obtained with the parameterization, which also includes the low-$\beta$ tail (see explanation in the text), are given by the dashed curves.} 
\label{fig:lqk}
\end{figure}


\begin{figure}
\epsfig{file=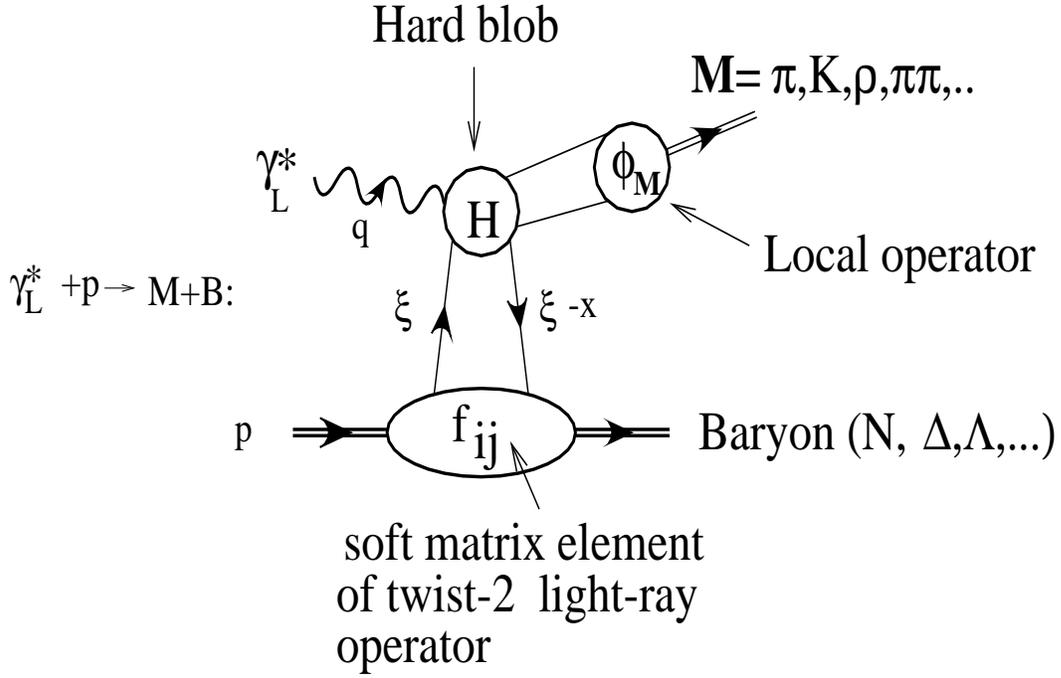,height=9cm,width=14cm}
\vspace{2cm}
\caption{The block structure of the amplitude for hard exclusive production of mesons $M$ by longitudinally polarized photons $\gamma_{L}^{\ast}$ in the reaction $\gamma_{L}^{\ast}+p \to M+B$, where $p$ is proton and $B$ is a baryon.}
\label{fig:panic}
\end{figure}

\newpage

\begin{figure}
\epsfig{file=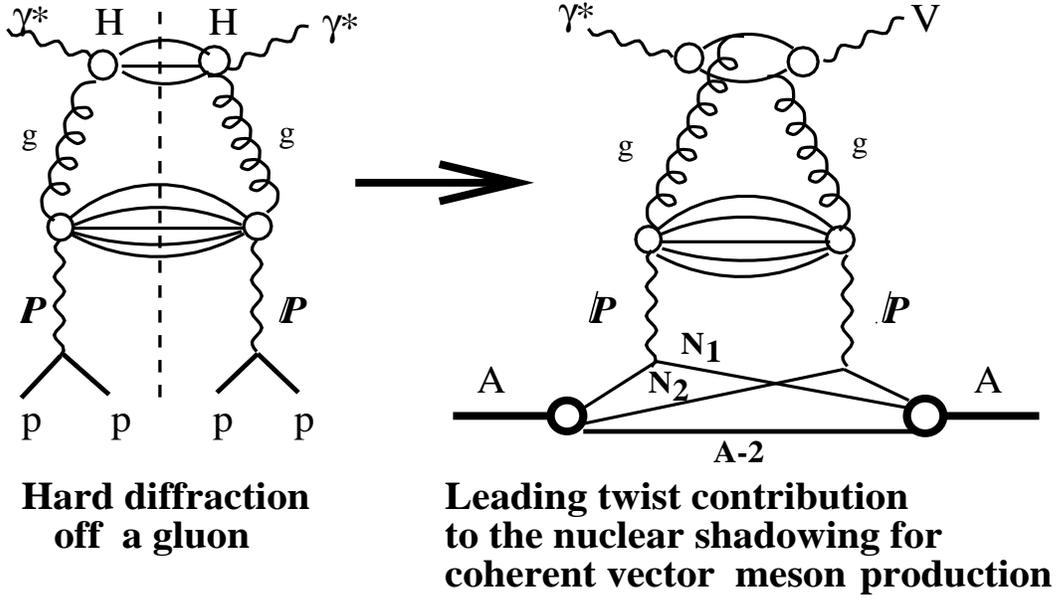,height=8cm,width=14cm}
\vspace{2cm}
\caption{Diagrams demonstrating the relationship between 
the gluon induced  hard diffraction on protons 
 and the leading twist contribution to nuclear shadowing in exclusive
 hard vector meson production in DIS on nuclei.}
\label{fig:shadrhobl}
\end{figure}


\begin{figure}
\epsfig{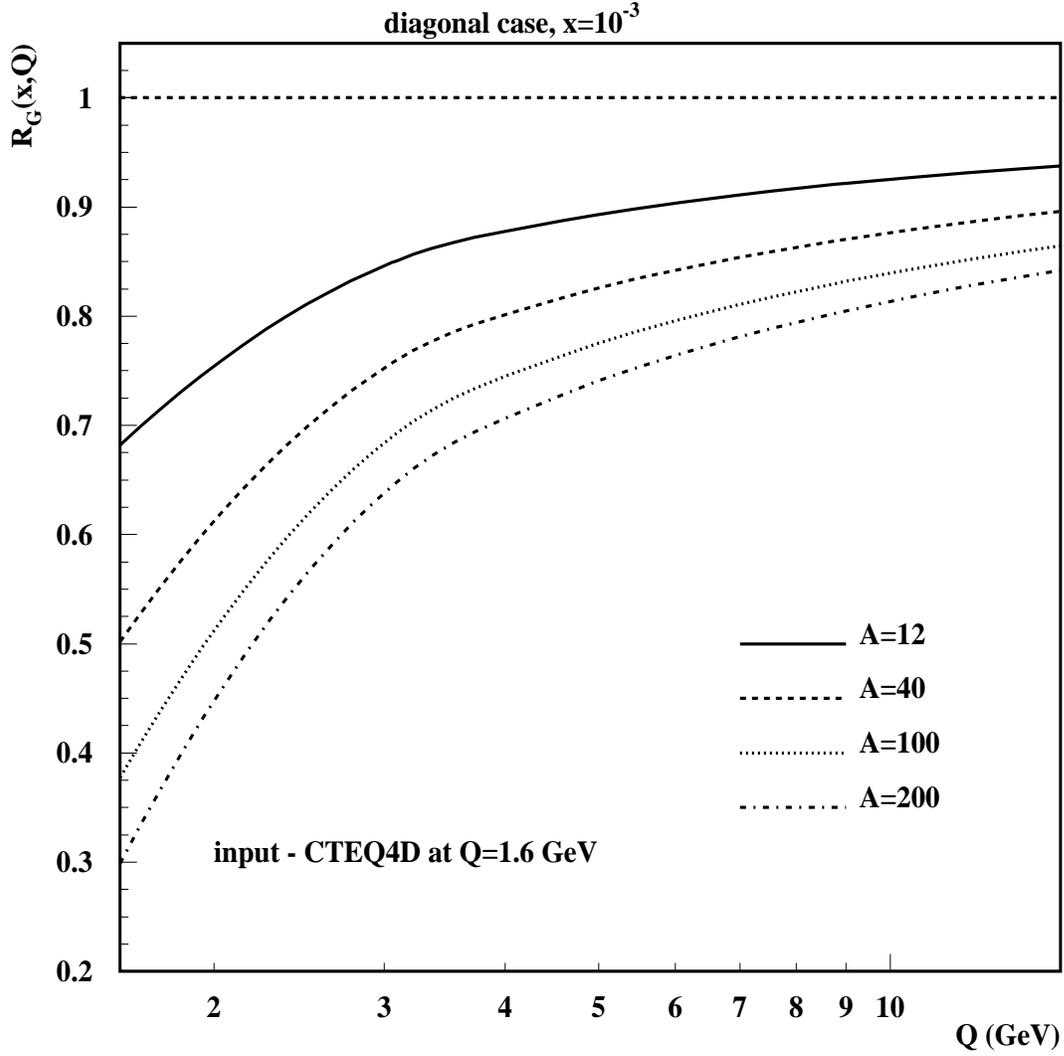}
\vspace{1cm}

\caption{The ratio of the conventional  nuclear to nucleon gluon parton densities $R_{G}(x,Q)$
 as a function of $Q$ at $x$=10$^{-3}$. The input distribution for the QCD evolution is CTEQ4D.}

\vspace{1cm}
\label{fig:ratq}
\end{figure}


\begin{figure}
\epsfig{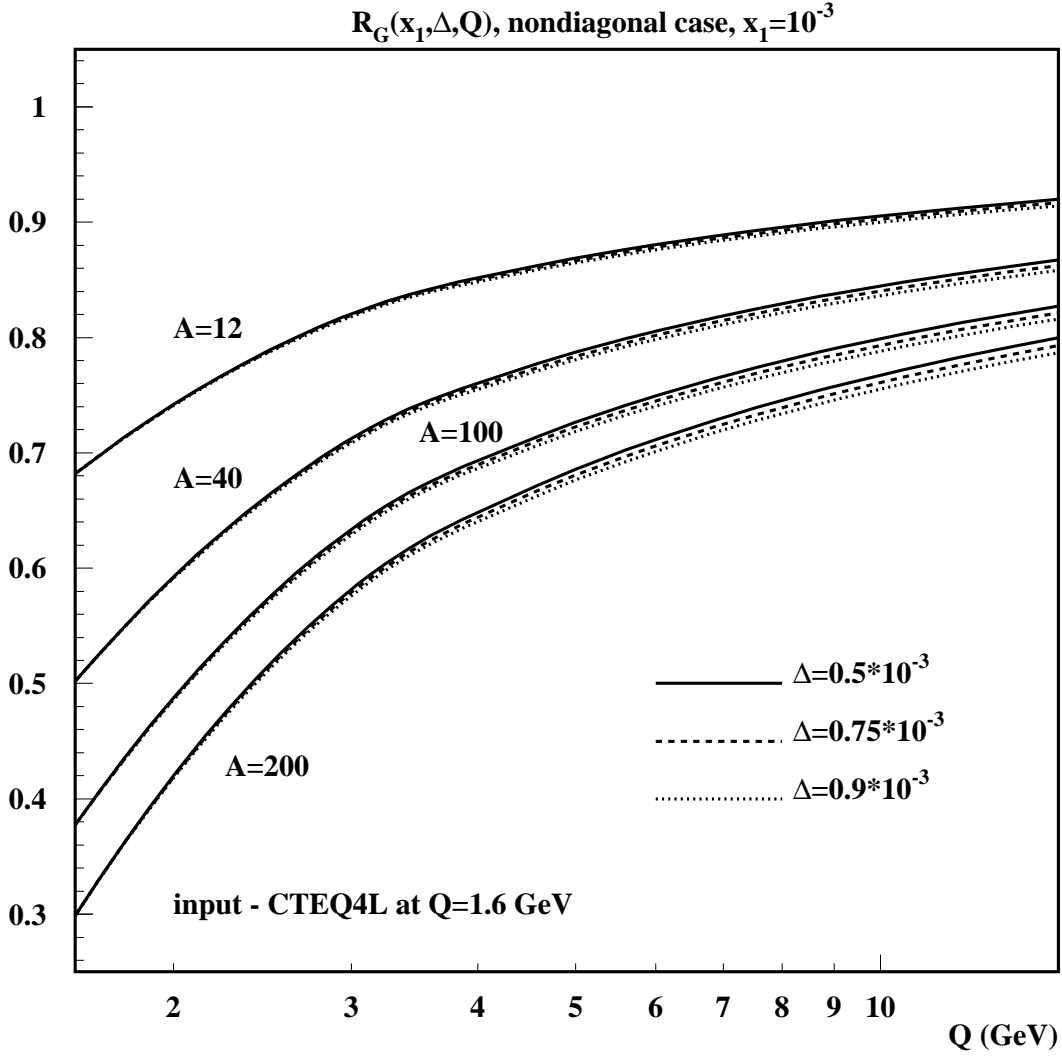}
\vspace{1cm}

\caption{The ratio of the skewed nuclear to nucleon gluon densities $R_{G}(x_1,x_2,Q)$ as a function of $Q$ at $x_1$=10$^{-3}$ and at $\Delta=0.5 \times 10^{-3}$, $\Delta=0.75 \times 10^{-3}$ and  $\Delta=0.9 \times 10^{-3}$. The three curves for each $\Delta$ are barely distinguishable. The input distribution for the QCD evolution is CTEQ4L.}

\label{fig:ratnd}
\end{figure}


\begin{figure}
\epsfig{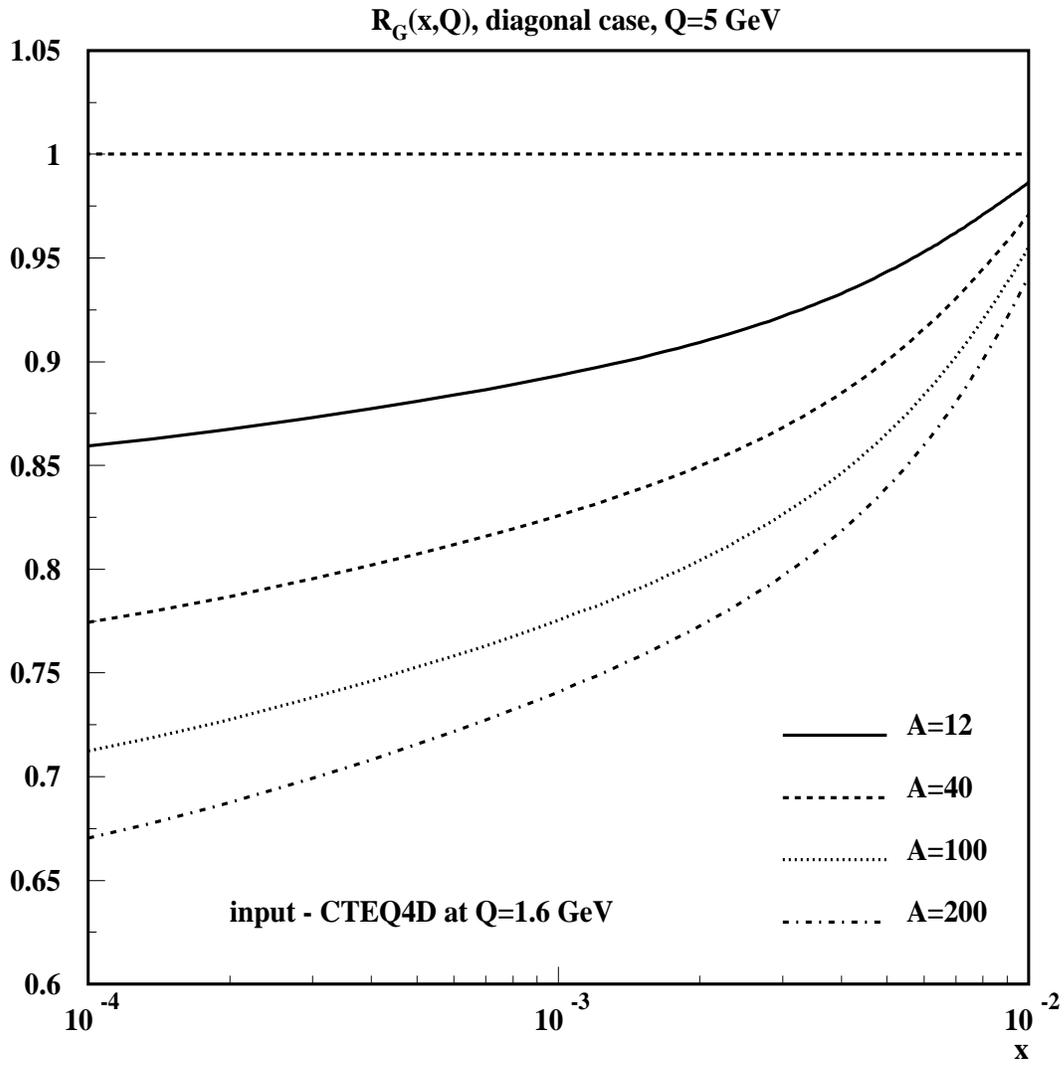}

\vspace{1cm}

\caption{The ratio of the conventional  nuclear to nucleon gluon parton densities $R_{G}(x,Q)$
 as a function of $x$ at $Q$=5 GeV. The input distribution for the QCD evolution  is CTEQ4D.}
\vspace{1cm}
\label{fig:ratx}
\end{figure}


\begin{figure}
\epsfig{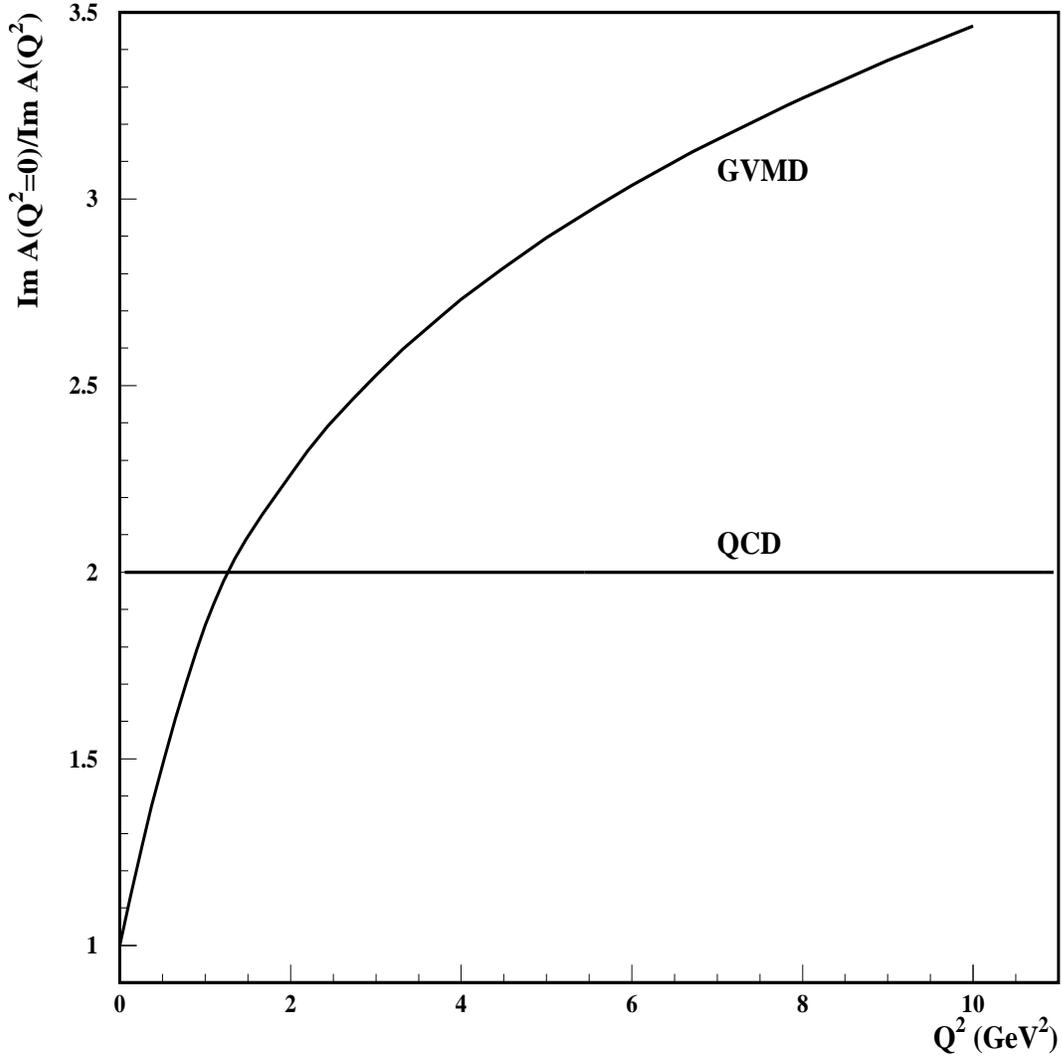}

\vspace{1cm}

\caption{The ratio of the imaginary parts of the amplitudes for real and virtual photon production by virtual photons as a function of $Q^2$. The straight line is the pQCD prediction [59]. The result of the calculation using the generalized vector meson dominance model (see Eq.\ (\ref{extra1}) is given by curve.}
\label{fracdvcs}
\end{figure}


\begin{figure}
\epsfig{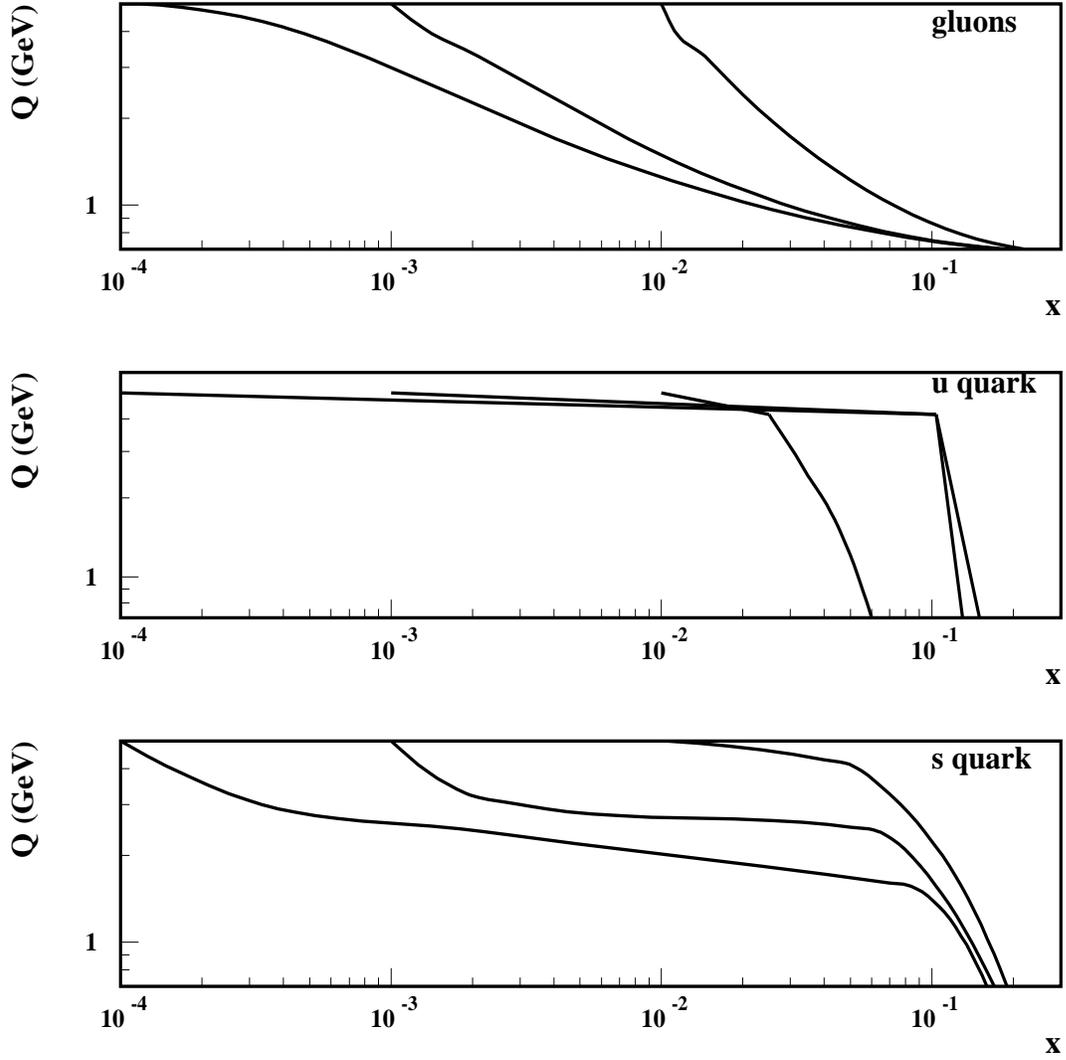}

\vspace{1cm}
\caption{The QCD evolution path in the $x-Q$-plane for the gluon, $u$ quark, and $s$ quark parton densities for the three final points, $x=10^{-4}$, $10^{-3}$, and $10^{-2}$. CTEQ4LQ with $Q_{0}$=0.7 GeV is used as an input for the QCD evolution.}

\label{fig:pathlq}
\end{figure}


\begin{figure}
\epsfig{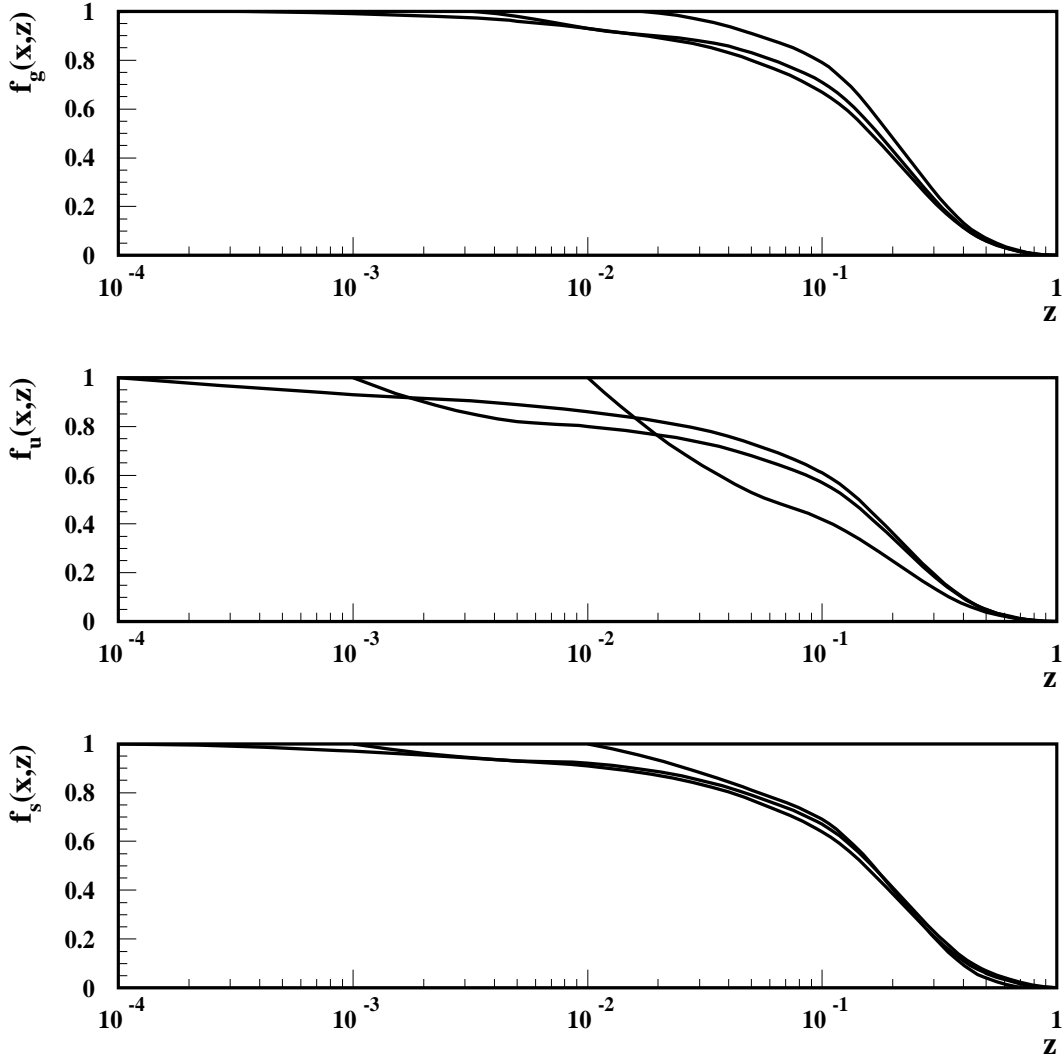}

\vspace{1cm}
\caption{Functions $f(x,z)$ as a function of $z$ and at $x=10^{-4}$, $10^{-3}$, and $10^{-2}$ for the gluon, $u$ quark, and $s$ quark parton densities. CTEQ4LQ with $Q_{0}$=0.7 GeV is used as an  input for the QCD evolution.}

\label{fig:displq}
\end{figure}


\begin{figure}
\epsfig{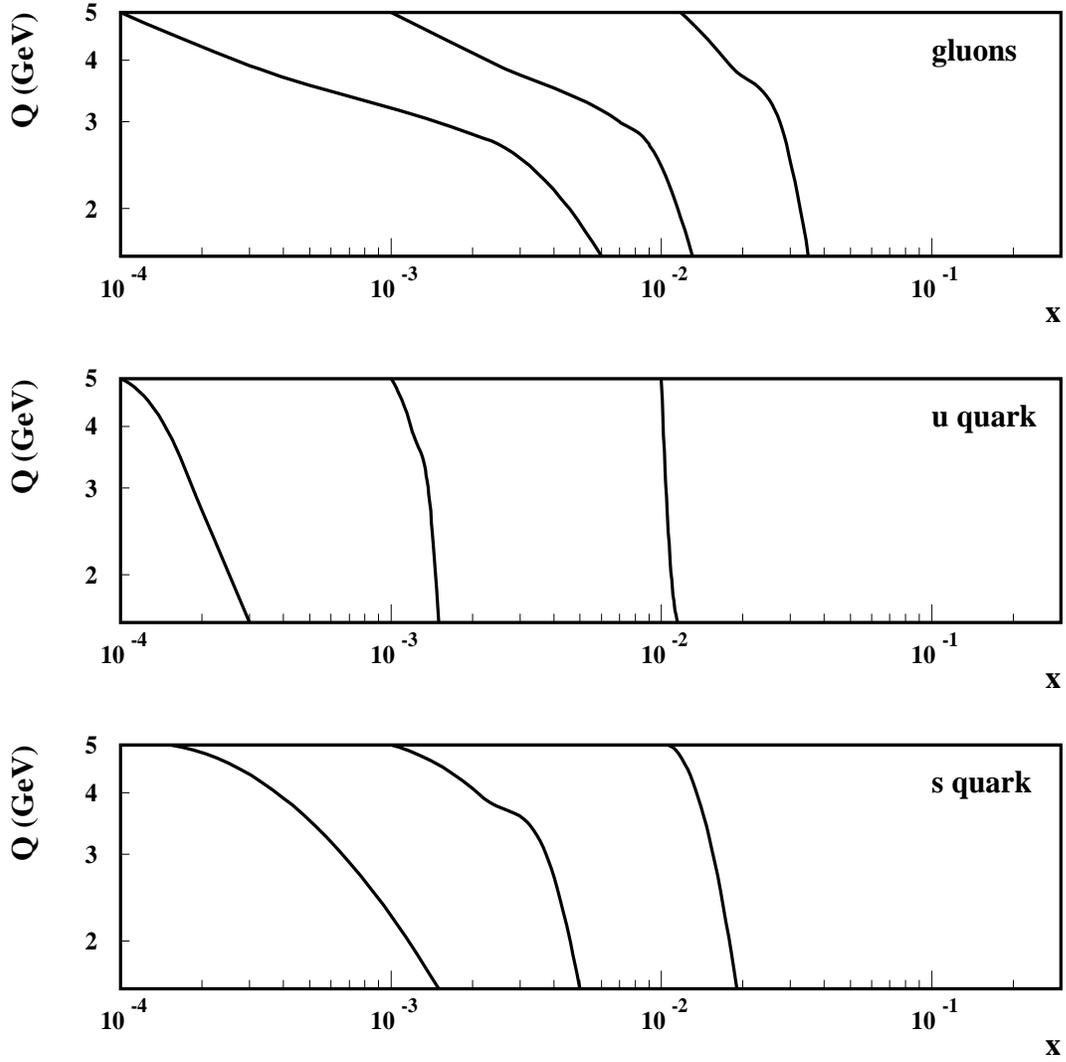}
\vspace{1cm}

\caption{The QCD evolution path in the $x-Q$-plane for the gluon, $u$ quark, and $s$ quark  parton densities for the three final points, $x=10^{-4}$, $10^{-3}$, and $10^{-2}$.
CTEQ4D with $Q_{0}$=1.6 GeV is used as an  input for the QCD evolution.}
\label{fig:path4d}
\end{figure}


\begin{figure}
\epsfig{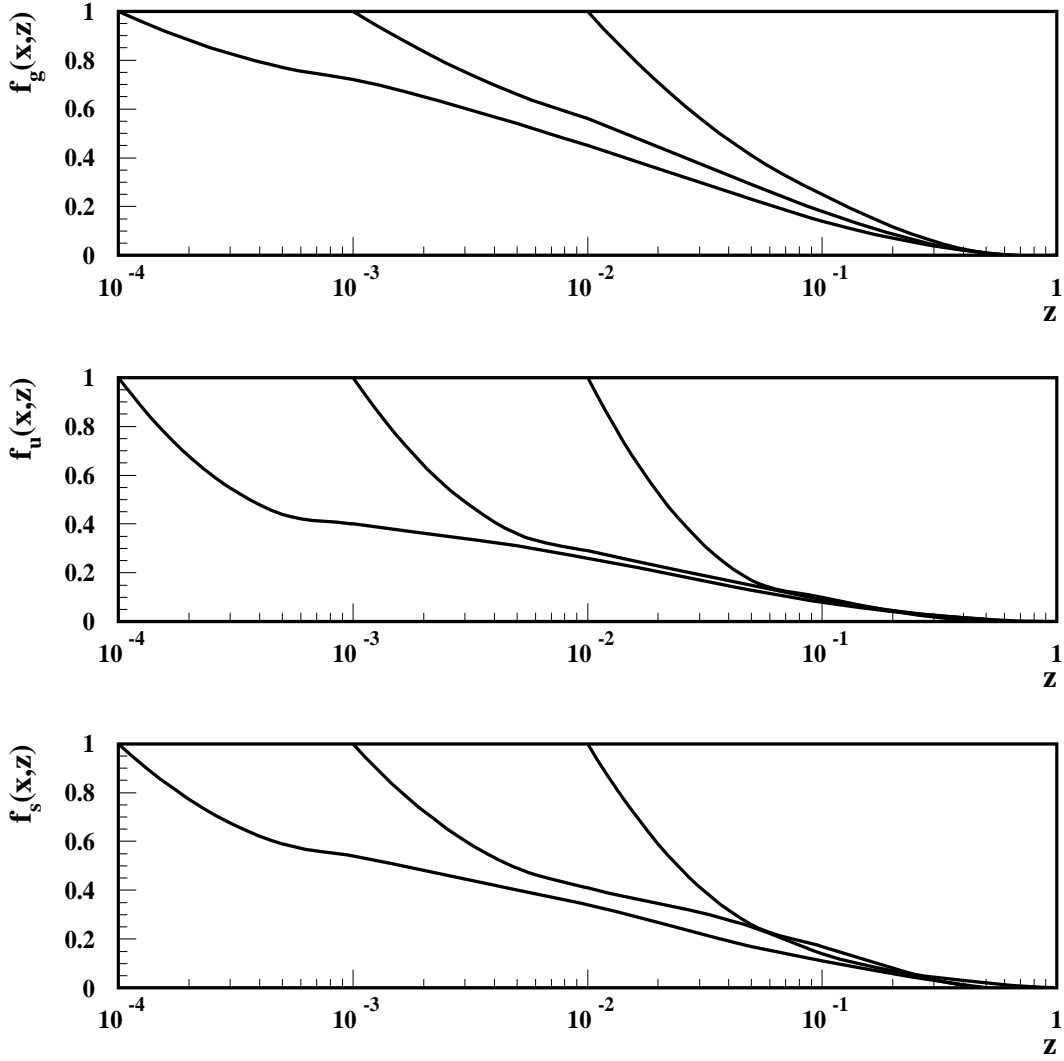}
\vspace{1cm}
\caption{Functions $f(x,z)$ as a function of $z$ and at $x=10^{-4}$, $10^{-3}$, and $10^{-2}$ for the gluon, $u$ quark, and $s$ quark parton densities. CTEQ4D with $Q_{0}$=1.6 GeV is used as an  input for the QCD evolution.}

\label{fig:disp4d}
\end{figure}


\begin{figure}

\epsfig{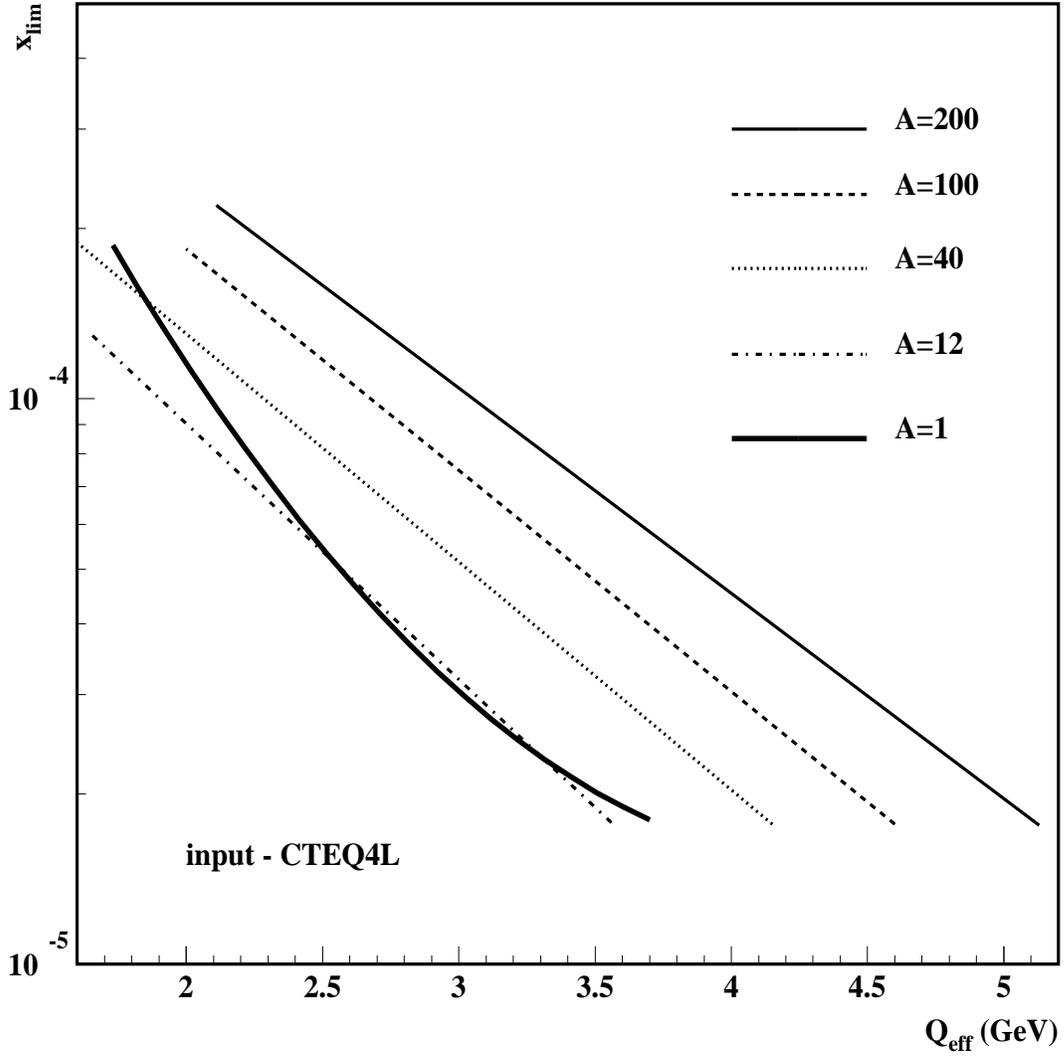}

\vspace{1cm}
\caption{The unitarity boundary for the inelastic $q\bar{q}$-nucleus 
cross sections for nuclei with $A$=12, 40, 100, and 200. The unitarity boundary for the inelastic $q\bar{q}$-nucleon cross section is presented as a thick solid curve.}

\label{bound1}
\end{figure}


\begin{figure}

\epsfig{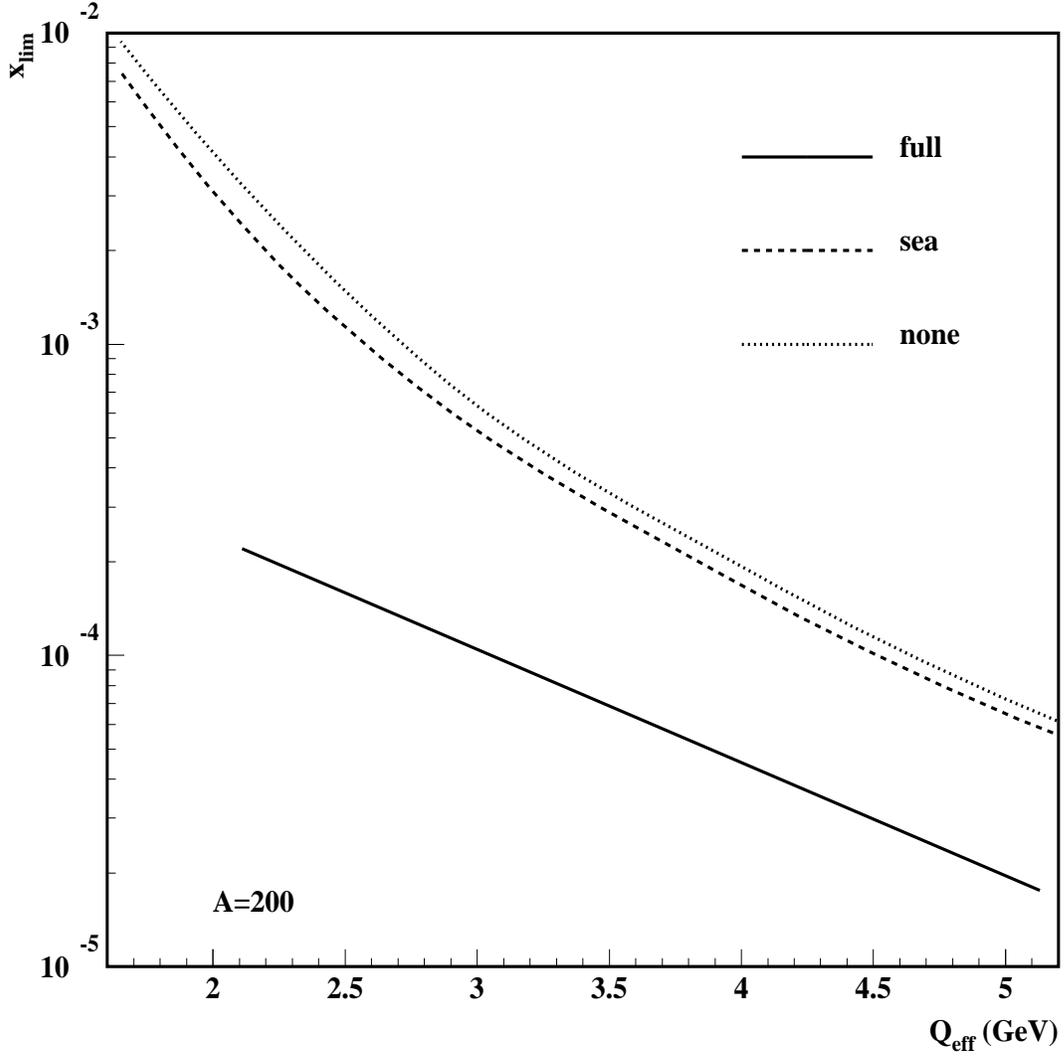}
\vspace{1cm}

\caption{The unitarity boundary for the inelastic $q\bar{q}$-nucleus 
cross sections for  nuclei with $A$=200. The amount of nuclear shadowing for gluons is varied: the solid curve corresponds to the full amount of shadowing, the dashed curve corresponds to the gluons shadowed as the sea quarks, and the dotted curve is for gluons without shadowing.}
\label{bound2}
\end{figure}


\begin{figure}

\epsfig{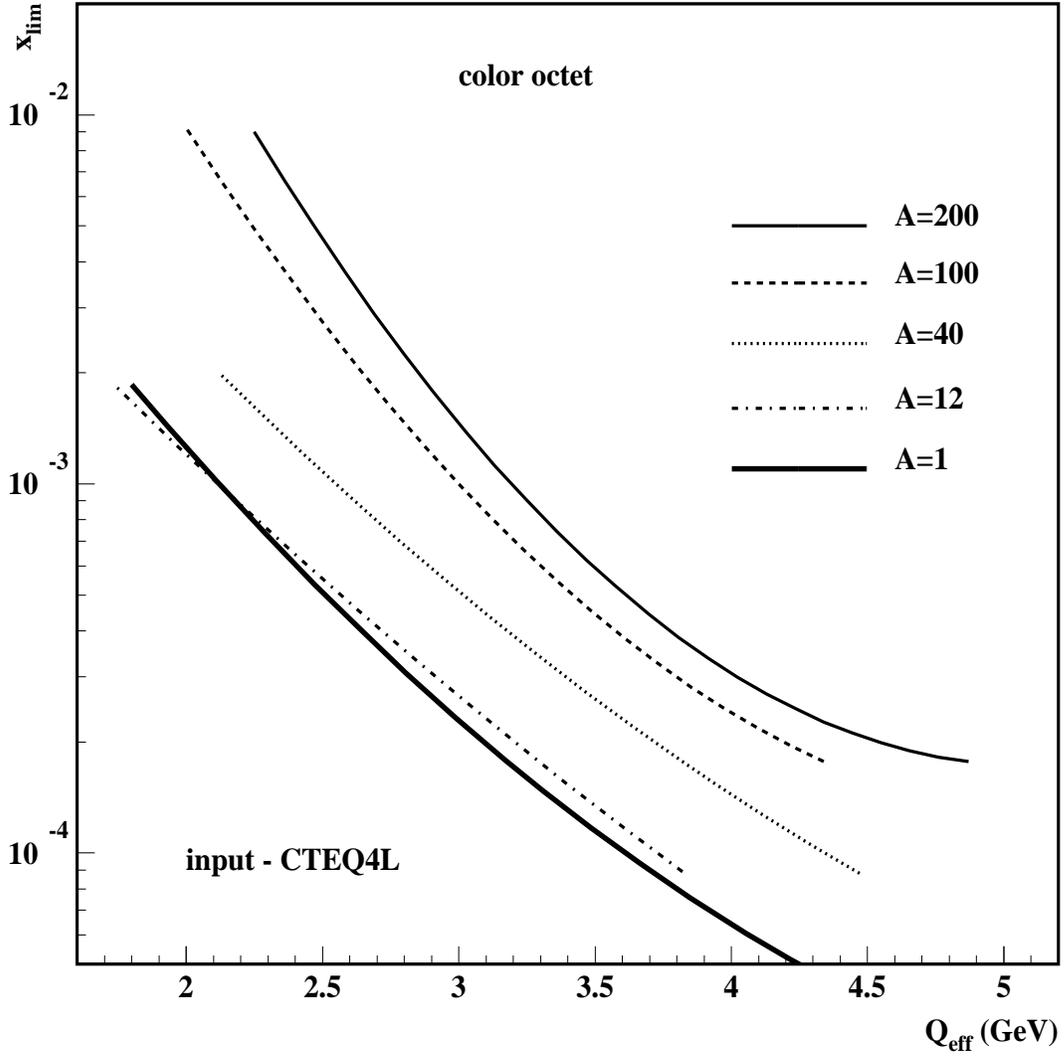}
\vspace{1cm}

\caption{The unitarity boundary for the inelastic $q\bar{q}g$ (color octet)-nucleus 
cross sections for nuclei with $A$=12, 40, 100, and 200. The unitarity boundary for the inelastic $q\bar{q}g$-nucleon cross section is presented as a thick solid curve.}

\label{bound3}
\end{figure}


\begin{figure}

\epsfig{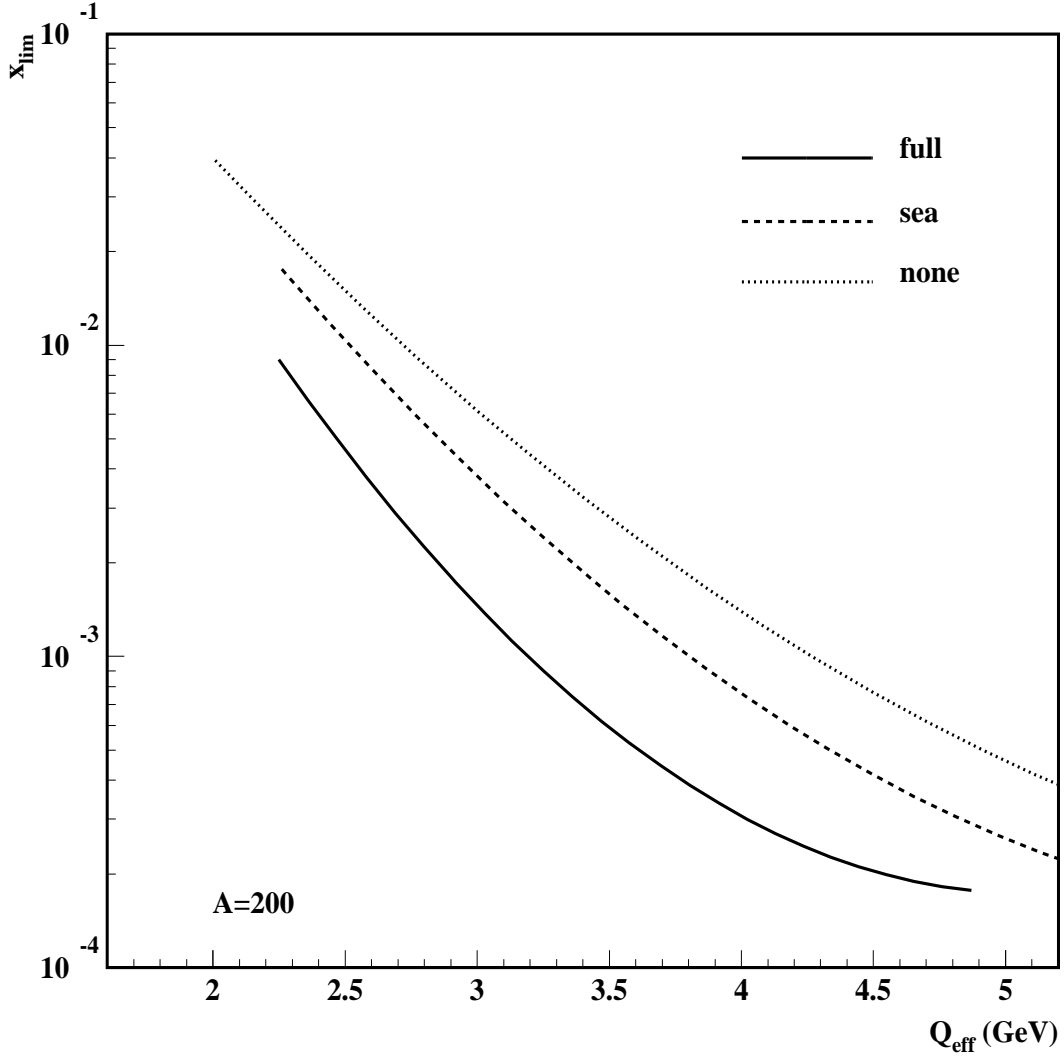}
\vspace{1cm}

\caption{The unitarity boundary for the inelastic $q\bar{q}g$ (color octet)-nucleus 
cross sections for  nuclei with $A$=200. The amount of nuclear shadowing for gluons is varied: the solid curve corresponds to the full amount of shadowing, the dashed curve corresponds to the gluons shadowed as the sea quarks, and the dotted curve is for gluons without shadowing.}
\label{bound4}
\end{figure}


\begin{figure}
\epsfig{file=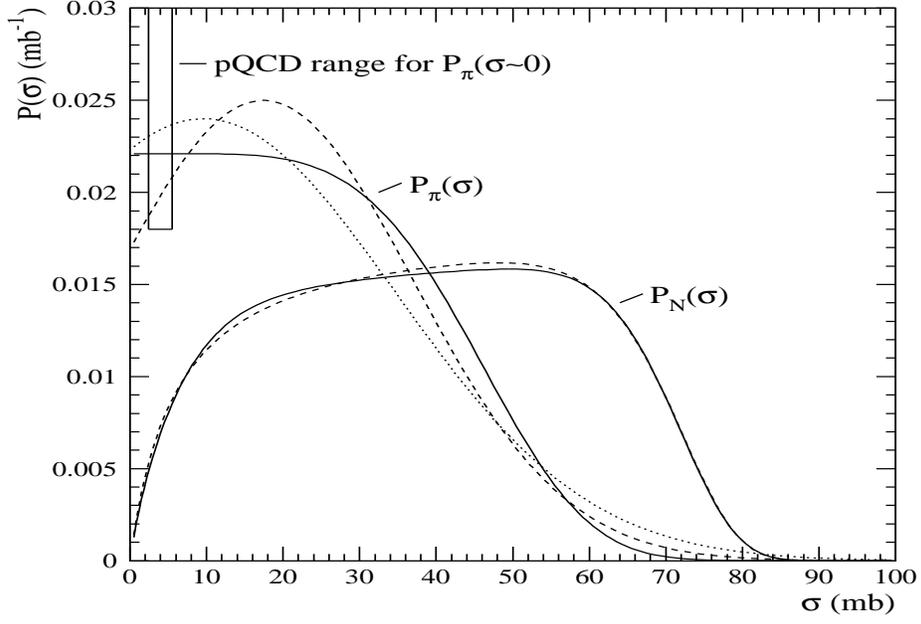,height=14cm,width=14cm}
\vspace{1cm}

\caption{The distribution over cross sections $P(\sigma)$ for protons and pions, presented in Eq.\ (\ref{param}).}
\label{fig4}
\end{figure}


\begin{figure}
\epsfig{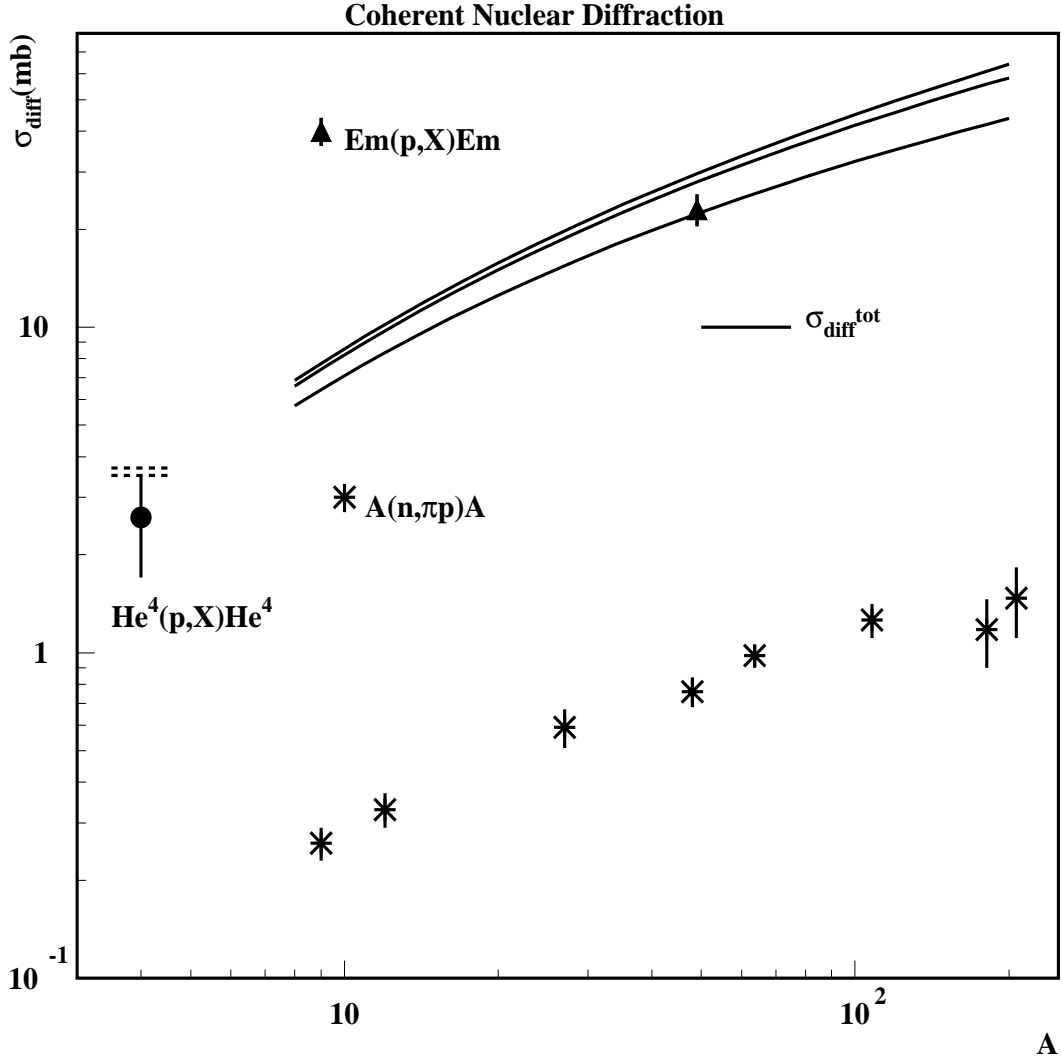}
\vspace{1cm}

\caption{The cross section of coherent diffraction dissociation  of protons and neutrons on nuclei  as a function of the atomic number  $A$. The solid curves are the theoretical prediction of Eq.\ (\ref{Adiff}). The data on the reaction $n+A \to p \pi^{-}+A$ [81]
  is presented as stars and the data for the emulsion targets [82] is presented as triangles. The data point, presented by the full circle, corresponds to $A=4$ ($^4$He) and is extracted from [77] (see Eq.\ (\ref{th2})). The theoretical prediction for coherent diffraction on  $^4$He is given by the dashed curves (see Eq.\ (\ref{th1})). 
}
\label{fig5}
\end{figure}

\end{document}